\documentclass[11pt,oneside]{article}
\usepackage[english]{babel}
\usepackage[utf8]{inputenc}
\usepackage[T1]{fontenc}
\usepackage{pmboxdraw}
\usepackage[letterpaper,top=1in,bottom=1in,left=1in,right=1in,marginparwidth=1in]{geometry}

\usepackage{amsmath,amsfonts,amssymb,amsthm}
\usepackage{mathtools}
\usepackage{graphicx}
\graphicspath{{./figs/}}
\usepackage{url}
\usepackage[colorlinks=true, allcolors=blue]{hyperref}
\usepackage{bbold}
\usepackage{xcolor}
\usepackage{float}
\usepackage{braket}
\usepackage{slashed}
\usepackage[bottom]{footmisc}
\usepackage{multirow}
\usepackage{booktabs}
\usepackage{caption}
\usepackage{cleveref}
\usepackage{tabularx}
\usepackage{longtable}
\usepackage{enumitem}
\usepackage{xspace}
\usepackage{setspace}
\usepackage{etoolbox}
\usepackage{listings}
\usepackage[parfill]{parskip}
\usepackage{cite}
\numberwithin{equation}{section}
\newcommand{\reportnumber}{FERMILAB-PUB-26-0208-T}
\usepackage[angle=0,scale=1,color=black,firstpage=true,opacity=1]{background}
\SetBgContents{\footnotesize\sf{\reportnumber}}
\SetBgPosition{current page.north east}
\SetBgHshift{-2in}
\SetBgVshift{-0.5in}

\usepackage{tikz}
\usetikzlibrary{decorations.pathmorphing,decorations.markings,arrows.meta,positioning}
\usepackage{xcolor}
\usepackage{alltt}

\tikzset{
  fermion/.style={draw, postaction={decorate},
    decoration={markings, mark=at position 0.55 with {\arrow{Latex}}}},
  antifermion/.style={draw, postaction={decorate},
    decoration={markings, mark=at position 0.55 with {\arrowreversed{Latex}}}},
  boson/.style={draw, decorate,
    decoration={snake, amplitude=1.5pt, segment length=5pt}},
  scalar/.style={draw, dashed},
  vertex/.style={circle, fill=black, inner sep=0pt, minimum size=4pt},
}
\usepackage[most]{tcolorbox}

\definecolor{AgentUserColor}{RGB}{180, 60, 60}
\definecolor{AgentResponseColor}{RGB}{120, 120, 255}
\definecolor{AgentSystemColor}{RGB}{180, 0, 180}
\definecolor{AgentToolColor}{RGB}{200, 100, 0}
\definecolor{AgentSubagentColor}{RGB}{40, 140, 70}

\newtcolorbox{agentuser}{%
  enhanced, breakable,
  colback=white, colframe=AgentUserColor, boxrule=0.6pt, arc=1.5mm,
  width=\linewidth, top=8pt, bottom=8pt, left=10pt, right=10pt,
  fontupper=\small,
  attach boxed title to top left={yshift=-6pt, xshift=3.5mm},
  boxed title style={colback=white, colframe=white, boxrule=0pt,
    left=0pt, right=0pt, top=3pt, bottom=0pt, sharp corners},
  coltitle=AgentUserColor, fonttitle=\sffamily,
  title={\raisebox{0pt}[0pt][0pt]{User}},
}

\newtcolorbox{agentresponse}{%
  enhanced, breakable,
  colback=white, colframe=AgentResponseColor, boxrule=0.6pt, arc=1.5mm,
  width=\linewidth, top=8pt, bottom=8pt, left=10pt, right=10pt,
  fontupper=\small,
  attach boxed title to top left={yshift=-6pt, xshift=3.5mm},
  boxed title style={colback=white, colframe=white, boxrule=0pt,
    left=0pt, right=0pt, top=3pt, bottom=0pt, sharp corners},
  coltitle=AgentResponseColor, fonttitle=\sffamily,
  title={\raisebox{0pt}[0pt][0pt]{Agent}},
}

\newtcolorbox{agentsystem}{%
  enhanced, breakable,
  colback=white, colframe=AgentSystemColor, boxrule=0.6pt, arc=1.5mm,
  width=\linewidth, top=8pt, bottom=8pt, left=10pt, right=10pt,
  fontupper=\small,
  attach boxed title to top left={yshift=-6pt, xshift=3.5mm},
  boxed title style={colback=white, colframe=white, boxrule=0pt,
    left=0pt, right=0pt, top=3pt, bottom=0pt, sharp corners},
  coltitle=AgentSystemColor, fonttitle=\sffamily,
  title={\raisebox{0pt}[0pt][0pt]{System}},
}

\newtcolorbox{agenttool}[1]{%
  enhanced, breakable,
  colback=white, colframe=AgentToolColor, boxrule=0.5pt, arc=1mm,
  width=\linewidth, left=1.5em, right=1em, top=6pt, bottom=6pt,
  fontupper=\small,
  attach boxed title to top left={yshift=-5pt, xshift=3.5mm},
  boxed title style={colback=white, colframe=white, boxrule=0pt,
    left=0pt, right=0pt, top=3pt, bottom=0pt, sharp corners},
  coltitle=AgentToolColor, fonttitle=\sffamily\small,
  title={\raisebox{0pt}[0pt][0pt]{#1}},
}

\newtcolorbox{agentsubagent}[1]{%
  enhanced, breakable,
  colback=white, colframe=AgentSubagentColor, boxrule=0.5pt, arc=1mm,
  width=\linewidth, left=1.5em, right=1em, top=6pt, bottom=6pt,
  fontupper=\small,
  attach boxed title to top left={yshift=-5pt, xshift=3.5mm},
  boxed title style={colback=white, colframe=white, boxrule=0pt,
    left=0pt, right=0pt, top=3pt, bottom=0pt, sharp corners},
  coltitle=AgentSubagentColor, fonttitle=\sffamily\small,
  title={\raisebox{0pt}[0pt][0pt]{Subagent: #1}},
}

\newtcolorbox{toolbox}[1]{%
  enhanced, breakable,
  colback=white, colframe=AgentToolColor, boxrule=0.5pt, arc=1mm,
  width=\linewidth, left=1.5em, right=1em, top=6pt, bottom=6pt,
  fontupper=\small,
  attach boxed title to top left={yshift=-5pt, xshift=3.5mm},
  boxed title style={colback=white, colframe=white, boxrule=0pt,
    left=0pt, right=0pt, top=3pt, bottom=0pt, sharp corners},
  coltitle=AgentToolColor, fonttitle=\sffamily\small,
  title={\raisebox{0pt}[0pt][0pt]{#1}},
}

\usepackage{mdframed}
\mdfdefinestyle{grayboxstyle}{%
  backgroundcolor=gray!5,
  topline=true,
  bottomline=true,
  leftline=false,
  rightline=false,
  innerleftmargin=10pt,
  innerrightmargin=10pt,
  innertopmargin=8pt,
  innerbottommargin=8pt,
  linewidth=0.8pt,
  linecolor=gray!60,
  leftmargin=0pt,
  rightmargin=0pt}
\newmdenv[style=grayboxstyle]{graybox}

\lstdefinestyle{pythonstyle}{
  language=Python,
  basicstyle=\small\ttfamily,
  keywordstyle=\color{blue},
  stringstyle=\color{red!70!black},
  commentstyle=\color{green!50!black},
  breaklines=true,
  frame=lines,
  rulecolor=\color{gray!60},
  backgroundcolor=\color{gray!5},
  numbers=none,
  xleftmargin=0.5em,
  xrightmargin=0.5em,
}
\lstdefinestyle{mathematicastyle}{
  basicstyle=\small\ttfamily,
  keywordstyle=\color{blue!70!black},
  commentstyle=\color{green!50!black},
  breaklines=true,
  frame=lines,
  rulecolor=\color{gray!60},
  backgroundcolor=\color{gray!5},
  numbers=none,
  xleftmargin=0.5em,
  xrightmargin=0.5em,
}
\lstdefinestyle{jsonstyle}{
  basicstyle=\small\ttfamily,
  stringstyle=\color{red!70!black},
  breaklines=true,
  frame=lines,
  rulecolor=\color{gray!60},
  backgroundcolor=\color{gray!5},
  numbers=none,
  xleftmargin=0.5em,
  xrightmargin=0.5em,
}

\newcommand{\pkgname}{HEPTAPOD\xspace}
\newcommand{\adname}{Diagrammatica\xspace}
\newcommand{\fc}{{FeynCalc}\xspace}
\newcommand{\fg}{{FeynGraph}\xspace}
\newcommand{\MCP}{MCP\xspace}
\newcommand{\EDA}{EDA\xspace}

\title{\textbf{Agentic Diagrammatica: Towards Autonomous Symbolic Computation in High Energy Physics}}

\author{Tony Menzo$^{1,2}$\footnote{Corresponding author: \texttt{amenzo@ua.edu}},\, Alexander Roman$^{1}$, George T. Fleming$^{2}$, \\
Sergei Gleyzer$^{1}$, Konstantin T.~Matchev$^{1}$, Stephen Mrenna$^{2}$ \\
{$^1$}\textit{Department of Physics and Astronomy, University of Alabama, Tuscaloosa, AL 35487, USA}\\
{$^2$}\textit{Fermi National Accelerator Laboratory, Batavia, IL 60510, USA}
}

\date{\today}

\begin{document}
\maketitle

\begin{center}
\begin{minipage}{13cm}
\begin{center}
{\Large\bfseries Abstract}
\end{center}
\vspace{0.1in}
We present \textit{\adname}, a symbolic computation extension to the HEPTAPOD agentic framework, which enables LLM agents to plan and execute multi-step theoretical calculations.
Symbolic computation poses a distinctive reliability challenge for LLM agents, as correctness is governed by implicit mathematical conventions that are not encoded in a form that can be easily checked in the computational backend.
We identify two complementary remedies, tool-constrained computation and targeted knowledge grounding, and pursue the first as the primary architecture.
Concretely, we concentrate the agent's action distribution onto tool calls with convention-fixing semantics, in which the agent specifies a compact, human-auditable diagram specification and a trusted backend performs the symbolic or numerical manipulations exactly.
The toolkit provides two complementary calculation paths consuming a shared diagram specification: Naive Dimensional Analysis (NDA) for order-of-magnitude rate estimates and Exact Diagrammatic Analysis (EDA) for tree-level symbolic calculations via automatic FeynCalc code generation, both supplemented by automatic Feynman diagram enumeration and a navigable theory knowledge base.
The architecture is validated on two benchmarks:
(1)~an exhaustive catalog of all tree-level, single-vertex $1\to 2$ partial decay widths across scalar, fermion, and vector parents, with complete massless and threshold limits and Standard Model validation; and
(2)~an NDA sensitivity study of the muon decay multiplicity $\mu \to \nu_\mu\bar{\nu}_e + n(e^+e^-) + e^-$, determining the maximum observable $n$ at current and planned muon experiments.

\end{minipage}
\end{center}
\vspace*{0.8cm}

\newpage
{\small\tableofcontents}
\newpage

\setcounter{page}{1}

\section{Introduction}
\label{sec:intro}

Symbolic computation has a long and distinguished tradition in high-energy physics (HEP), supported by a rich ecosystem of computer algebra systems (CAS) and specialized packages.
Dedicated systems such as REDUCE~\cite{REDUCE}, FORM~\cite{FORM}, Cadabra~\cite{Cadabra}, and Redberry~\cite{Redberry} provide HEP-oriented symbolic engines, while general-purpose platforms like Mathematica~\cite{Mathematica} and SymPy~\cite{SymPy} host widely used packages including \fc~\cite{FeynCalc1,FeynCalc2}, FeynArts/FormCalc~\cite{FeynArts,FormCalc}, Package-X~\cite{PackageX,PackageX2}, and FIRE~\cite{FIRE}. 
These tools are powerful, but even routine calculations are manual, multi-step processes.   For example, computing tree-level decay widths or leading-order cross sections requires: constructing an amplitude from Feynman rules, squaring it, performing spin sums, evaluating Dirac traces, substituting kinematics, and extracting a partial width.
Many steps require familiarity with both the physics and the idiosyncrasies of the specific software package being used. This is why hands-on tutorials are now routinely scheduled at graduate-level summer schools and workshops \cite{MC4BSM}.

Large language models (LLMs) have recently emerged as general-purpose computational interfaces capable of planning and executing multi-step workflows through structured tool invocation~\cite{LLMAgentSurvey,AIScience}.
In our previous work, we introduced the HEP Toolkit for Agentic Planning/Programming, Orchestration, and Deployment (\pkgname) framework~\cite{Menzo:2025cim}, demonstrating how an LLM agent can orchestrate Monte Carlo simulation pipelines through schema-validated tool invocation.
Since then, a number of papers employing agentic LLMs have appeared that target different stages of the HEP workflow, including event generation~\cite{MadAgents}, collider analysis pipelines~\cite{CoLLM,ColliderAgent}, experimental data analysis~\cite{HEPDataAnalysisAgents,Moreno:2026mqk}, experimental design and detector simulation~\cite{GRACE}, theory hypothesis generation and reinterpretation~\cite{Agrawal:2026fermiacc}, lattice QCD automation~\cite{Tan:2026physmaster}, and broader multi-domain science~\cite{ArgoLOOM}.
In all of these systems the LLM orchestrates domain-specific computational engines rather than replacing them, providing the procedural knowledge that connects individual steps into coherent workflows.

This paper introduces \textit{\adname}\footnote{From the Latinized \emph{diagrammatica}: ``the systematic art of diagrams'' and paying homage to M.~Veltman's book~\cite{Veltman:1994} of the same name.}, a symbolic computation extension to the \pkgname framework that enables LLM agents to plan and execute multi-step theoretical calculations.
Symbolic computation poses a distinctive challenge compared to the Monte Carlo orchestration demonstrated in~\cite{Menzo:2025cim}.
Frontier LLMs are already capable of impressive symbolic manipulation (solving integrals, simplifying expressions, writing working computer algebra code), but \emph{reliability}, not capability, is the bottleneck.
The multi-step calculations typical of HEP phenomenology require dozens of symbolic operations chained together under a shared set of implicit conventions that must remain internally consistent throughout.
Autoregressive generation provides no guarantee that this consistency will be maintained across an extended workflow, even when each individual step is well within the model's capabilities.
An LLM navigating this space without external structure can produce plausible but silently wrong results, including dropped terms, misapplied identities, and inconsistent conventions. 
Even purpose-built transformer models trained to perform specific symbolic subtasks, such as predicting squared amplitudes directly from Feynman diagrams~\cite{Alnuqaydan:2022ncd}, can achieve high but imperfect accuracy, illustrating the difficulty of replacing traditional CAS tools for exact symbolic computation. 
Additionally, auditing free-form symbolic code for such errors is \textit{at least as hard} as writing it from scratch.
In this paper we identify two complementary remedies to this reliability problem.
The first is \emph{tool-constrained computation}: instead of asking the agent to generate the symbolic calculation directly, the action space is restricted to schema-validated tool calls that fix conventions by construction, where the agent specifies a compact, human-auditable diagram specification and a trusted backend performs the algebra exactly.
The second is \emph{targeted knowledge grounding}: delivering domain-specific conventions and worked examples to the agent at the moment of each critical decision, rather than loading reference material in bulk.
Tool constraints provide structural reliability that is independent of the model's knowledge, but bounds the agent's capabilities to the tool designer's foresight; knowledge grounding preserves the full action space but offers only probabilistic suppression of errors.
This paper primarily pursues the tool-constrained path, realized through a shared diagram specification and deterministic symbolic tools, while using a navigable theory knowledge base as a complementary layer (\cref{sec:tools_vs_knowledge}).

This approach contrasts with the predominant paradigm in the growing literature on LLM agents in HEP~\cite{LLMAgentSurvey,AIScience,MadAgents,CoLLM,ColliderAgent,HEPDataAnalysisAgents,GRACE,Agrawal:2026fermiacc,Tan:2026physmaster,ArgoLOOM,Moreno:2026mqk}, in which the agent writes free-form scripts or shell commands executed against existing software, with the reliability burden falling entirely on the model's ability to produce correct code from context.
The key distinction of the approach pursued here (as well as in Ref.~\cite{Menzo:2025cim}) is to design tool schemas that reduce the agent's action space to a physically meaningful set of choices, fixing domain conventions by construction and providing a more natural computational interface for the LLM.
This yields two concrete advantages: \emph{reliability}, as correctness of the mechanical steps becomes a property of the interface rather than of the model, eliminating entire classes of silent errors independently of model capability, and \emph{interpretability}, as the agent's decisions are compressed into a small number of human-readable fields that a domain expert can audit at a glance rather than reviewing pages of generated code where errors may be invisible without re-deriving the calculation from scratch.

The toolkit provides two complementary calculation paths at different fidelity levels, both consuming a shared diagram specification:
\begin{enumerate}[leftmargin=2em]
    \item \textbf{NDA (Naive Dimensional Analysis):} Order-of-magnitude estimates of decay widths, cross sections, and branching ratios using dimensional analysis, phase space volumes, and coupling power counting.
    \item \textbf{EDA (Exact Diagrammatic Analysis):} Complete tree-level calculations via automated \fc~\cite{FeynCalc1,FeynCalc2} code generation.
    Produces symbolic expressions, \LaTeX{} output, and Python-callable numerical evaluations from a single diagram specification.
\end{enumerate}
This two-fidelity structure mirrors how phenomenologists naturally work, allowing for order-of-magnitude estimates to guide the decision to commit to a full calculation.
It also extends the toolkit's reach beyond what either path could achieve alone, since exact analytic results are restricted to low-multiplicity final states where phase space integration is tractable, while NDA provides useful rate estimates for arbitrarily complex processes.
Both paths consume a shared, LLM-compatible, convention-fixing diagram specification, introduced in this work, and are further supplemented by automatic Feynman diagram enumeration via \fg~\cite{FeynGraph} as well as a navigable theory knowledge base. 

We define two benchmark tasks in this paper to serve as both architecture validation and as reproducible test cases for other agentic platforms.
\textbf{Task~1} (\cref{sec:catalog}) is an exhaustive catalog of all tree-level, single-vertex $1\to 2$ decay widths across scalar, fermion, and vector parents, validated against Standard Model partial widths.
\textbf{Task~2} (\cref{sec:muon}) is an NDA-based sensitivity study of $\mu \to \nu_\mu\bar{\nu}_e + n(e^+e^-) + e^-$ as a function of multiplicity $n$, stressing diagram enumeration and rate estimation across a large variety of diagrams.
Both tasks are presented through curated, annotated agent transcripts from single Claude Code \cite{Claude} sessions, demonstrating the full workflow from task prompt to compiled deliverables without manual intervention.

The remainder of this paper is organized as follows.
\Cref{sec:reliability} establishes the case for tool-constrained agentic calculation through an entropy decomposition of the agent's action distribution, with a token-level illustration for $Z\to e^+e^-$.
\Cref{sec:architecture} presents the toolkit architecture: the shared diagram specification, the NDA and \EDA calculation paths, the \fg enumeration engine, and the theory knowledge base.
The architecture is validated on two benchmarks: \cref{sec:catalog} presents the $1\to 2$ decay rate catalog (Task~1), exercising the \EDA path, and \cref{sec:muon} presents the muon decay sensitivity study (Task~2), exercising the NDA path.
\Cref{sec:conclusions} discusses implications and future directions.

\section{Agentic symbolic programming}

\subsection{Reliability, uncertainty, and the case for tool-constrained calculation}
\label{sec:reliability}

The primitive outputs of an LLM agent, operating at decision step $t$ with context $C_t$, are sequentially and recursively generated token sequences $x_{t+1:t+m}$ of varying length $m$ over some \emph{vocabulary} $\mathcal{V}$, sampled from a $\theta$-parameterized autoregressive distribution 
\begin{equation}
    p_\theta(x_{t+1:t+m} \mid C_t)
\end{equation}
representing the probability of the next token sequence $x_{t+1:t+m}$, given the previous context $C_t$. In practice, the actual quantity of interest isn't the raw token sequence but the \emph{action}, $\mathcal{A}$, it implements \emph{i.e.}~the code block it produces, the natural-language instruction it conveys, \emph{etc}. 
Agentic programming fundamentally operates on this action abstraction. 
Many distinct `microscopic' token sequences can implement the same `macroscopic' action, so a given action naturally forms an equivalence class whose probability is obtained by marginalizing over all implementing sequences of any length $m$,
\begin{equation}
    p(\mathcal{A}_t \mid C_t)
    \;=\;
    \sum_{m=1}^{\infty}\;
    \sum_{\{x_{t+1:t+m}\,\mapsto\, a\}}
    p_\theta(x_{t+1:t+m} \mid C_t).
    \label{eq:action_prob}
\end{equation}
A particularly important action in agentic programming is the \emph{tool call}. 
A tool is an external computational backend---a function, script, or service---exposed to the agent in the initialization context $C_0$ through a convention-dependent schema specifying its name, inputs, and return types. A tool call is then a token sequence that forms a structured JSON object matching this schema, which the inference serving infrastructure can intercept and route to that backend (typically the user's machine). 
Importantly, adding or removing tools doesn't change the model's token vocabulary, but for particular tasks it can significantly reshape the conditional distribution of \cref{eq:action_prob} by concentrating probability mass onto the structured subset of the action space corresponding to valid tool invocations. 
Some inference providers enforce this concentration even more strongly through constrained decoding~\cite{Willard:2023}, in which tokens incompatible with the tool schema are masked to zero probability at each generation step.

The practical value of concentrating $p(\mathcal{A}_t \mid C_t)$ depends on the task. 
The obvious candidates are exact arithmetic, external state access, and formal grammar adherence. 
More subtle, and more relevant here, are tasks where the correctness criteria are implicit — nowhere encoded in a form the computational backend can check against reliably. 
Symbolic computation occupies this category. 
Unlike compiled code, where type errors and syntax violations halt execution and provide immediate feedback, a computer algebra system has no equivalent system to check its input against. Correctness is defined \textit{implicitly} by the mathematical conventions of the domain, which are external to the system itself. 
An LLM navigating this space without external structure can only rely on statistical correlations in its training data, correlations that are themselves unreliable, since conventions in the literature are not always unanimous.

A high-energy physics symbolic calculation is a particularly acute instance, where conventions are numerous, interrelated, and drawn from a literature in which different sources make incompatible choices. 
A correct calculation may have to simultaneously respect consistent choices for the metric signature ($+{-}{-}{-}$ vs.\ $-{+}{+}{+}$), spinor normalization ($\bar u u = 2m$ vs.\ $\bar u u = 1$), the sign of the covariant derivative, the normalization of $\mathrm{SU}(N)$ generators ($\operatorname{Tr}[T^a T^b] = \delta^{ab}/2$ vs.\ $\delta^{ab}$), the choice of gauge, and the phase conventions in CKM and PMNS matrices, among others.
An LLM trained on the full literature will have seen all of these conventions represented, often without explicit declaration, and the statistical correlations it relies on are precisely the wrong tool for enforcing internal consistency across a multi-step calculation where every convention choice must be made once and propagated uniformly.

Trusted symbolic computation with LLM agents therefore requires that the action space be restricted to tools where a fixed set of conventions is encoded in the computational backend, or that additional context be provided to the agent fixing the conventions explicitly.
Developing tools that provide these semantic constraints reliably for common HEP computations is the main focus of this work.
The simplest way to realize this is to alter what inputs the agent operates on: instead of asking the model to realize the calculation directly in free-form symbolic text, one asks it to specify a typed action whose arguments correspond to physically meaningful choices---vertex class, external spins, topology, mass assignments, and other convention-fixing data---and the backend performs the algebra exactly.
In this design, the LLM is responsible for selecting the correct calculation, while the symbolic engine is responsible for executing it.
A similar principle underlies the reliability of LLM-based theorem proving in Lean~\cite{LeanDojo}, where probability mass is concentrated on valid proof tactics verified against a structured proof state rather than on free-form natural-language reasoning.
The resulting \emph{verification asymmetry} shifts the locus of human oversight from a high-dimensional verification problem to a low-dimensional one.  It is easier to audit a compact, human-readable diagram specification with a few physically meaningful fields than a full generated script.

To organize the case for tool-constrained calculation quantitatively, it is useful to decompose the uncertainty of the agent's action distribution, measured by the conditional Shannon entropy
\begin{equation}
    \Delta_\mathcal{A}(C_L)
    \;\equiv\;
    H(\mathcal{A}_t \mid C_L)
    \;=\;
    -\sum_{a} p(a \mid C_L)\,\log_2 p(a \mid C_L),
    \label{eq:entropy_def}
\end{equation}
where $L$ denotes the amount of context provided, into three components with qualitatively distinct origins and architectural remedies:
\begin{equation}
    \Delta_{\mathcal{A}}(C_L)
    \;\approx\;
    \underbrace{\Delta_T(C_0)}_{\substack{\text{task}\\\text{uncertainty}}}
    \;+\;
    \underbrace{\Delta_C(L)}_{\substack{\text{context}\\\text{uncertainty}}}
    \;+\;
    \underbrace{\Delta_E}_{\substack{\text{execution}\\\text{uncertainty}}}.
    \label{eq:entropy_decomposition}
\end{equation}
This decomposition is approximate (ignoring possible correlations between components, for example) but it identifies three distinct failure modes of free-form agentic generation, each mapping onto a concrete design decision in the toolkit.

\begin{figure}[t!]
\centering
\includegraphics[width=0.85\textwidth]{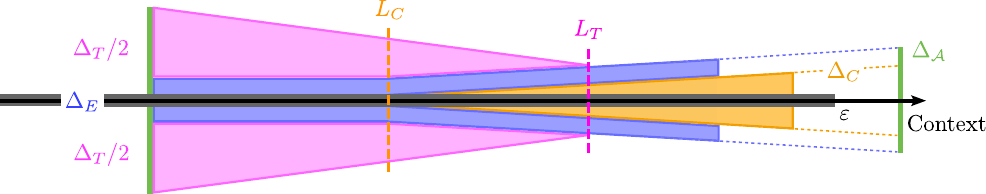}
\caption{Schematic depiction of each action uncertainty (entropy) component as a function of context length. 
}
\label{fig:entropy_triangle}
\end{figure}

\begin{itemize}[leftmargin=2em]

    \item \textbf{Task uncertainty} $\Delta_T(C_0)$ captures ambiguity in \emph{what to compute}.
    $\Delta_T$ decreases as relevant task-specifying context (\emph{e.g.}~the process definition, particle identities, vertex types, desired output format) is provided, and reaches approximately zero at a natural scale $L_T$, the minimum token count needed to fully specify the calculation.
    Below $L_T$ the agent is genuinely uncertain about what it is computing; above $L_T$, additional task-specification context yields diminishing returns.

    \item \textbf{Execution uncertainty} $\Delta_E$ is the variability introduced by generating the calculation as unstructured token-by-token output.
    $\Delta_E$ is largely \emph{independent of how much documentation is in the context}.
    Reference material loaded in bulk does not suppress token-level variability at critical positions (function names, index placements, sign choices, convention-dependent prefactors) because the agent must still retrieve and apply the right convention at the right moment from a long context, so $\Delta_E$ appears as a roughly constant offset in the $\Delta$ vs.\ $L$ diagram of \cref{fig:entropy_triangle}.
    The most direct remedy is to constrain the action space~\cite{Toolformer,ReAct}, replacing free-form generation with a schema-validated tool call that limits every feasible completion to a physically meaningful choice and eliminates the silent-error modes described above.
    A qualitatively different mechanism is to deliver targeted domain knowledge \emph{on demand at the moment of the critical decision}, rather than loading it in bulk; this can partially suppress $\Delta_E$ and is discussed in \cref{sec:tools_vs_knowledge}.

    \item \textbf{Context uncertainty} $\Delta_C(L)$ is the degradation of attention and retrieval fidelity as context length grows~\cite{lostinthemiddle}.
    Unlike $\Delta_T$, this term \emph{increases} with $L$, introducing a fundamental tension in which adding context to resolve task ambiguity simultaneously inflates context-induced errors.
    There is a natural scale $L_C$ marking the onset of this degradation.
    When $L_T \ll L_C$, the task is fully resolved before degradation sets in and reliable operation is possible.
    When $L_T \gtrsim L_C$, the documentation needed to resolve the task pushes the context past the degradation onset, and the workflow becomes unreliable regardless of how much reference material is provided.
    The remedy is to increase the information density of each context token by structuring domain knowledge as compact, retrievable nodes rather than raw long-form text~\cite{RAG}, exposing exactly the information the agent needs in a small number of tokens and shifting $L_C$ to larger values.
    Crucially, when such a node is retrieved \emph{on demand at the moment} the agent faces a convention-dependent choice, it not only avoids inflating $\Delta_C$ but also partially suppresses $\Delta_E$ by resolving the specific ambiguity at the point where execution variability originates.
    It is the \emph{timing of delivery}, not the content, that distinguishes this from bulk documentation (see \cref{sec:tools_vs_knowledge}).

\end{itemize}

Finally, it is reasonable to expect an irreducible floor $\varepsilon > 0$ in the action entropy of any finite autoregressive model. Even with perfect context and a fully constrained action space, the stochastic sampling procedure and finite model capacity prevent the conditional distribution from collapsing to a delta function on the correct action.
Whether such a floor can be made rigorous, and how it scales with model size, is an open question; for the present purposes it suffices to note that the three architectural remedies above---structured specifications, tools, and skills---are designed to reduce the \emph{controllable} components of $\Delta$ as far as possible, leaving whatever irreducible residual remains as the dominant source of uncertainty.

\subsection{A token-level illustration}
\label{sec:token_entropy}

The suppression of each uncertainty component by the corresponding architectural remedy can be made explicit at the level of individual token predictions.
Consider an agent that receives the prompt:

\begin{agentuser}
Compute the tree-level partial width for a scalar particle $S$ decaying to a fermion--antifermion pair via a Yukawa interaction $\mathcal{L} \supset \bar\psi (g_S + g_P\gamma_5)\psi\phi$.
Retain general masses and both scalar and pseudoscalar couplings.
\end{agentuser}

\noindent
For a well-posed prompt like the one above, the task uncertainty $\Delta_T$ is already small, the parent particle, daughters, vertex type, and desired output are all explicitly stated.
How small depends on the model's pretraining corpus and resulting familiarity with the domain, a model without QFT in its training data would have high $\Delta_T$ regardless of how precise the prompt is.
A correct response using \fc~\cite{FeynCalc1,FeynCalc2} requires the following six steps, common to every $1\to 2$ decay supported by the toolkit.

\begin{enumerate}[leftmargin=2em]

    \item \textbf{Write the invariant amplitude.}
    For a general scalar--pseudoscalar Yukawa vertex,
    \begin{equation}
        i\mathcal{M}
        =
        \bar{u}(p_1)\,i(g_S + g_P\gamma_5)\,v(p_2).
    \end{equation}
    This is the \emph{only} process-specific ingredient; all subsequent steps are determined by it.

    \item \textbf{Form the spin-summed squared amplitude.}
    Square $\mathcal{M}$, sum over final-state spins, and apply the completeness relations $\sum_s u\bar{u} = \slashed{p}_1 + m_1$ and $\sum_s v\bar{v} = \slashed{p}_2 - m_2$.
    No initial-state spin averaging is needed for a scalar parent.

    \item \textbf{Evaluate the Dirac trace.}
    The spin sum produces a single trace,
    \begin{equation}
        \sum_{\text{spins}} |\mathcal{M}|^2
        =
        \operatorname{Tr}\!\bigl[
            (\slashed{p}_1 + m_1)(g_S + g_P\gamma_5)
            (\slashed{p}_2 - m_2)(g_S - g_P\gamma_5)
        \bigr],
    \end{equation}
    which separates into $g_S^2$ and $g_P^2$ contributions; mixed $g_S g_P$ terms vanish by the trace of an odd number of gamma matrices.

    \item \textbf{Apply two-body kinematics.}
    All scalar products reduce to masses: $p_1 \cdot p_2 = (M^2 - m_1^2 - m_2^2)/2$.

    \item \textbf{Assemble the partial width.}
    \begin{equation}
        \Gamma
        =
        \frac{|\vec{p}\,|}{8\pi M^2}
        \sum_{\text{spins}} |\mathcal{M}|^2,
        \qquad
        |\vec{p}\,|
        =
        \frac{\lambda^{1/2}(M^2, m_1^2, m_2^2)}{2M},
    \end{equation}
    where $\lambda(a,b,c) = a^2 + b^2 + c^2 - 2ab - 2bc - 2ca$ is the K\"all\'en function.

    \item \textbf{Simplify and extract limits.}
    For equal daughter masses $m_1 = m_2 = m_f$,
    \begin{equation}
        \Gamma(S \to f\bar{f})
        =
        \frac{M}{8\pi}\sqrt{1 - \frac{4m_f^2}{M^2}}
        \left[
            g_S^2\!\left(1 - \frac{4m_f^2}{M^2}\right)
            + g_P^2
        \right].
        \label{eq:scalar_ff}
    \end{equation}
    The massless limit gives $\Gamma = (g_S^2 + g_P^2)\,M/(8\pi)$.

\end{enumerate}

Steps~2--6 are purely mechanical, determined entirely by the vertex structure and external-state quantum numbers chosen in step~1, yet generating them as free-form \fc code requires the agent to make approximately $N_c \sim 15$ critical token-level decisions: the relative sign in the $v$-spinor completeness relation ($\slashed{p} - m$, not $\slashed{p} + m$), the sign of $\gamma_5$ in the conjugate vertex, the correct \fc function for trace evaluation, the kinematic substitution rule for $p_1 \cdot p_2$, and the phase-space prefactor with the correct power of $M$, among others.
At each such position, different token choices produce physically distinct amplitudes so the token-level entropy at critical positions is a reasonable proxy for the action-level entropy of \cref{eq:action_prob}.
The joint entropy of the full output sequence satisfies the chain rule $H(x_1, \ldots, x_m) = \sum_i H(x_i \mid x_{<i})$, but the conditional entropy is near zero at the majority of positions, where the next token is determined by syntax and code structure.
The sum is dominated by the $N_c$ critical positions where convention-dependent choices arise, precisely because these are the positions at which the training corpus contains multiple incompatible conventions and the model's conditional distribution reflects this ambiguity.
Under the simplifying assumption that critical-token errors are approximately independent, which should be reasonable when they occur in distinct algebraic steps, the total execution entropy is bounded by $N_c \cdot h$, where $h$ is the local conditional entropy at each critical position.
At each such position, $h \sim \mathcal{O}(\text{few})$~bits, set by the effective vocabulary of $n \sim 10$--$30$ plausible alternatives that carry non-negligible probability mass after the softmax\footnote{The effective vocabulary $n$ at a given position refers not to the full token vocabulary $|\mathcal{V}|$, which is typically $\mathcal{O}(10^5)$, but to the number of tokens with logit values within a few units of the maximum. The remaining tokens contribute negligibly to the conditional entropy.}.
This gives an upper bound on the total execution entropy,
\begin{equation}
    \Delta_E^{\text{free-form}}
    \;\lesssim\;
    N_c \cdot h
    \;\sim\; 45\text{--}75~\text{bits}.
    \label{eq:free_form_entropy}
\end{equation}
A more direct measure of the practical impact is the success probability. Assuming a per-position error rate of $\epsilon \sim 0.05$, the probability that all $N_c$ critical tokens are correct is $(1-\epsilon)^{N_c} \approx 0.46$.
The conclusion is insensitive to the precise value of $\epsilon$, even at $\epsilon \sim 0.02$, the success rate is only $\sim\!74\%$, a failure rate that compounds rapidly when sweeping systematically over many processes.
The situation worsens if the agent resolves \fc conventions by ingesting raw documentation, where the relevant formula is buried among thousands of tokens~\cite{lostinthemiddle}, inflating $\Delta_C$ simultaneously.

Schema-validated tools reduce the viable token vocabulary at two levels.
First, the tool definitions injected into $C_0$ make a structured tool invocation the natural next action given the context, concentrating probability mass away from free-form symbolic code generation entirely.
Second, the tool call schema constrains each argument to a small enumerated set of valid field values.
Much like integrating out heavy degrees of freedom in an effective field theory collapses the path integral onto a smaller set of dynamical variables, the combined effect suppresses the next-token probability for the vast majority of $|\mathcal{V}|$, leaving an effective vocabulary of $n' \sim 10$ tokens per decision point.
The agent specifies only $N_c' \sim 3$--$5$ such fields (parent spin, daughter spins, vertex type), each drawn from an enumerated set of $n' \sim 10$ valid values.
The local conditional entropy per field is $h' = \log_2 n' \approx 3.3$~bits, so the total execution entropy is bounded by
\begin{equation}
    \Delta_E^{\text{tool}}
    \;\lesssim\;
    N_c' \cdot h'
    \;\sim\; 10\text{--}17~\text{bits}.
    \label{eq:tool_entropy}
\end{equation}
Crucially, all of this residual uncertainty is concentrated in a human-readable specification that a domain expert can audit at a glance, while a targeted skill retrieval of $\sim\!200$ tokens replaces the raw documentation, suppressing $\Delta_C$ and partially concentrating the distribution at convention-dependent decision points.
The compression is not a consequence of a more capable model; it comes from shrinking the space of actions the model can take, so that even a mediocre distribution over it has low entropy.
This observation motivates the architecture described in \cref{sec:architecture}: a shared structured specification that minimizes $\Delta_T$, schema-validated tools that compress $\Delta_E$, and a navigable theory knowledge base that suppresses $\Delta_C$ while also partially addressing $\Delta_E$ through targeted knowledge injection (\cref{sec:tools_vs_knowledge}).

\subsection{Tool constraints versus knowledge grounding}
\label{sec:tools_vs_knowledge}

The execution uncertainty $\Delta_E$ identified in the previous section admits two complementary remedies that, despite both involving the delivery of information to the agent, operate through qualitatively different channels and offer fundamentally different tradeoffs.

The first mechanism, emphasized in this paper, is \emph{tool-constrained computation}.
Instead of asking the agent to generate the symbolic calculation, the action space is restricted to schema-validated tool calls that fix conventions by construction.
The agent specifies \emph{what} to compute through a diagram specification, and a trusted backend determines \emph{how} to compute it by performing the algebra exactly.
This drives $\Delta_E$ to near zero \emph{structurally}, since correctness is a consequence of the interface design rather than the model's knowledge, and holds regardless of what context the agent has seen.
The cost is that the agent's capabilities are bounded by the tool designer's foresight.
It can compute any decay width the tools support, but it cannot notice that two seemingly unrelated results satisfy an unexpected identity, propose a novel parameterization, or derive a relation that the tool vocabulary does not anticipate.
The tools define the ceiling.

The second mechanism is \emph{knowledge-grounded computation}.
As discussed in \cref{sec:reliability}, bulk reference material loaded into context does not suppress $\Delta_E$ because the agent must still retrieve and apply the right convention at the right moment from a long context, which is essentially a $\Delta_C$ problem.
What distinguishes \emph{targeted} knowledge injection is timing.
A skills-graph node retrieved on demand at the exact moment the agent faces a convention-dependent choice directly concentrates the token distribution at that position, resolving the specific ambiguity that drives execution variability.
This is not ``more context'' in the sense of the $\Delta_C$ analysis but rather context that acts at the \emph{locus} of $\Delta_E$.
The distinction from task-specifying context, which reduces $\Delta_T$ by clarifying \emph{what} to compute, is equally important.
Targeted knowledge addresses \emph{how} to execute a step that is already specified, not what the step should be.
The result is a partial, probabilistic suppression of $\Delta_E$, not the structural guarantee of tool constraints, but a genuine reduction that scales with the quality and specificity of the injected knowledge.
The advantage is that the full action space remains available.
A knowledge-grounded agent could in principle recognize patterns across calculations, derive new identities from first principles, or explore parameterizations that the tool designer never considered.
The ceiling is set by the model's reasoning capability, not by the tool vocabulary.

In practice, the choice between these mechanisms is not binary but a spectrum, and the optimal design point depends on the nature of the task.
Mechanical steps where creativity adds no value and errors are catastrophic (Dirac traces, spin sums, polarization averaging, phase space integration) should be tool-constrained.
Strategic steps where judgment and exploration are valuable (which process to study, what limits to take, how to organize a survey of many processes) benefit from free-form reasoning grounded in domain knowledge.
The \adname toolkit already makes this split implicitly: the agent reasons freely about strategy while using tools for tactical execution.
Naming it explicitly as a design principle clarifies both the strengths and the limitations of the current architecture.

The two approaches are also complementary in practice, since an agent that explores freely with knowledge grounding can confirm its results via \EDA tool calls, combining the discovery potential of the first approach with the reliability guarantees of the second.
This suggests a natural trajectory for future agentic systems, beginning with the tool-constrained architecture presented here and progressively expanding the boundary of free-form reasoning as the knowledge base deepens and verification tools mature.

\section{Architecture and design}
\label{sec:architecture}

The uncertainty decomposition of \cref{sec:reliability} identifies three architectural requirements for a reliable agentic symbolic toolkit: a compact structured representation that the agent can construct with low task uncertainty $\Delta_T$, a skills layer that supplies targeted domain knowledge to suppress context uncertainty $\Delta_C$, and schema-validated deterministic tools that drive execution uncertainty $\Delta_E$ toward zero for the mechanical steps of the calculation.
The six-step calculation procedure introduced in \cref{sec:token_entropy} for $S \to f\bar{f}$ is universal, since every $1\to 2$ decay follows the same sequence.
Step~1, writing the invariant amplitude, is the only process-specific ingredient.
Steps~2--6 are identical for every supported topology and are automated entirely by the \textbf{\textsf{ComputeSymbolicAmplitude}} tool (see \cref{fig:flowchart_task1}), which maps a structured diagram specification to a complete, self-contained \fc script.
This division, where the agent selects the physics and the backend executes the algebra, is the concrete realization of the $\Delta_E$ suppression argued for in \cref{sec:reliability}.
The key architectural enabler is a shared \emph{Diagram} specification (\cref{sec:diagram_spec}), a JSON dataclass consumed by all toolkit components that provides a common structured representation of the physics across stages.

Beyond the three entropy-motivated requirements, the architecture is organized around two additional design principles:
\begin{enumerate}[leftmargin=2em]
    \item \textbf{Persistent artifacts and provenance.}
    Generated scripts, parsed result sidecars, and exported Python functions are saved as concrete artifacts in the agent's session sandbox rather than discarded as transient outputs.
    Downstream tools consume results by reference, enabling auditable multi-step pipelines without manual transcription of symbolic expressions.

    \item \textbf{Multi-fidelity pathways.}
    The toolkit provides two calculational paths at different fidelity levels, both consuming the same diagram specification.
    NDA (\cref{sec:nda}) produces order-of-magnitude decay width and cross section estimates from dimensional analysis, phase space volumes, and coupling power counting, requiring no external software and applicable to arbitrary $n$-body final states.
    The \EDA path (\cref{sec:diagrammatica}) produces exact tree-level symbolic results via automatic \fc code generation.
    The current scope of the \EDA path is restricted to $1\to 2$ decays and $2\to 2$ scattering processes, where the phase space integration is analytic and the matrix element dependence on final-state kinematics drops out (for $1\to 2$) or reduces to Mandelstam variables (for $2\to 2$).
    Extension to $n \geq 3$ final-state particles, where phase space integration over non-trivial matrix elements is required, is left to future work.
    For such processes, where exact analytic phase space integration is generally not available, numerical Monte Carlo tools such as MadGraph~\cite{MadGraph}, already integrated into the \pkgname framework~\cite{Menzo:2025cim}, become the primary calculational path.
    MadGraph also serves as an independent numerical cross-check on the \EDA and NDA results for simpler processes.
    Both paths can be fed by automatic Feynman diagram enumerations engines, here we use \fg~\cite{FeynGraph} (\cref{sec:feyngraph}). 
    Agreement in scaling and order of magnitude between paths provides an internal consistency check; disagreement identifies where assumptions, topology assignments, or kinematics require inspection.
\end{enumerate}

The remainder of this section describes each component in turn.

\subsection{The LLM-compatible diagram specification}
\label{sec:diagram_spec}

The central architectural element of \adname is an LLM-compatible Feynman diagram format.
A single structured JSON object specifies a decay or scattering process and serves simultaneously as the input to all toolkit components and as a natural object for an LLM agent to construct, inspect, and manipulate.

Existing HEP software represents processes through formats designed for traditional programmatic workflows.
FeynArts~\cite{FeynArts} uses deeply nested Mathematica objects (\texttt{TopologyList}, \texttt{FeynmanGraph}) tied to the Mathematica ecosystem, and parton-level-generators like  CompHEP~\cite{Pukhov:1999gg}, CalcHEP~\cite{Belyaev:2012qa}, and MadGraph~\cite{MadGraph} specify processes through a domain-specific command syntax (for example, \texttt{generate p p > t t\~{} QCD<=2}).
These formats are expressive within their respective ecosystems, but none provides a single structured object that simultaneously serves as input to tools at different fidelity levels \emph{and} is natural for an LLM to construct under the token-entropy constraints of \cref{sec:token_entropy}.
The \adname diagram specification is designed for agentic manipulation, with a small, fixed vocabulary of field names, schema-validated enumerated types for vertex structures, human-readable particle labels, and a flat JSON structure requiring no ecosystem-specific knowledge.

\colorlet{initcol}{blue!70!black}
\colorlet{finalcol}{red!70!black}
\colorlet{propcol}{green!50!black}
\tikzset{
  fermion/.style={draw=#1, very thick, postaction={decorate},
    decoration={markings, mark=at position 0.6 with {\arrow[#1, scale=1.2]{Stealth}}}},
  fermion/.default=black,
  boson/.style={draw=#1, very thick, decorate,
    decoration={snake, amplitude=1.5pt, segment length=5pt}},
  boson/.default=black,
  scalar/.style={draw=#1, very thick, dashed},
  scalar/.default=black,
  vtx/.style={circle, fill=black, inner sep=0pt, minimum size=5pt},
}

\newsavebox{\numjsonbox}
\begin{lrbox}{\numjsonbox}%
\begin{minipage}[c]{0.68\textwidth}%
\begin{lstlisting}[style=jsonstyle]
{
  "initial":[{"label":"H","spin":0,
              "mass":125.2}],
  "final":[{"label":"b","spin":"1/2",
            "mass":2.79},
           {"label":"bbar","spin":"1/2",
            "mass":2.79}],
  "vertices":[{"type":"yukawa","coupling":"yb"}],
  "couplings":{"yb":0.017},
  "color_factor":3
}
\end{lstlisting}
\end{minipage}%
\end{lrbox}

\newsavebox{\symjsonbox}
\begin{lrbox}{\symjsonbox}%
\begin{minipage}[c]{0.60\textwidth}
\begin{lstlisting}[style=jsonstyle]
{
  "initial":[{"label":"e-","spin":"1/2"},
             {"label":"e+","spin":"1/2"}],
  "final":[{"label":"mu-","spin":"1/2"},
           {"label":"mu+","spin":"1/2"}],
  "vertices":[{"type":"vector-axial"} x2],
  "propagators":[{"label":"Z","spin":1,
                  "channel":"s"}]
}
\end{lstlisting}
\end{minipage}
\end{lrbox}

The \adname \texttt{Diagram} data structure has two modes.
Both share the same top-level schema with fields for initial- and final-state particles (label, spin, and optionally mass), vertices (Lorentz-structure type and coupling), propagators, color factors, and optional topology hints.
The vertex type field takes values from the enumerated set in \cref{tab:vertex_types}, a vocabulary of 20 physically distinct Lorentz structures with additional aliases.
Because every feasible completion maps to a valid physics choice, the schema-validated interface eliminates the silent-error modes of free-form code generation identified in \cref{sec:token_entropy}.

\begin{enumerate}[leftmargin=2em]

\item \textbf{Numerical (NDA) mode} (\texttt{Diagram}).
All masses and couplings are given explicit floating-point values.
The NDA path consumes this mode directly for immediate rate estimation, and \fg populates it automatically from the SM particle database during diagram enumeration.

\smallskip
\noindent\hspace*{-\leftmargin}\makebox[\textwidth][c]{%
\begin{minipage}[c]{0.22\textwidth}
\centering
\resizebox{0.9\linewidth}{!}{%
\begin{tikzpicture}
  \node[vtx] (v) at (0,0) {};
  \draw[scalar, line width=1.6pt] (-1.6,0) node[left,font=\Large] {$H$} -- (v);
  \draw[fermion, line width=1.6pt] (v) -- ( 1.5, 0.9) node[right,font=\Large] {$b$};
  \draw[fermion, line width=1.6pt] ( 1.5,-0.9) node[right,font=\Large] {$\bar{b}$} -- (v);
\end{tikzpicture}}
\end{minipage}%
\begin{minipage}[c]{0.06\textwidth}
\centering{\large$\longrightarrow$}
\end{minipage}%
\usebox{\numjsonbox}}

\item \textbf{Symbolic (EDA) mode} (\texttt{SymbolicDiagram}).
Masses and couplings are left as unresolved parameters, with only particle labels, spins, and vertex types specified.
The \EDA path uses this mode to derive general analytic formulas that can later be evaluated at any numerical values via the \textbf{\textsf{ConvertToPython}} tool (\cref{sec:diagrammatica}).

\smallskip
\noindent\hspace*{-\leftmargin}\makebox[\textwidth][c]{%
\begin{minipage}[c]{0.30\textwidth}
\centering
\resizebox{0.9\linewidth}{!}{%
\begin{tikzpicture}
  \node[vtx] (v1) at (-0.8, 0) {};
  \node[vtx] (v2) at ( 0.8, 0) {};
  \draw[fermion] (-1.5, 0.9) node[left] {$e^-$} -- (v1);
  \draw[fermion] (v1) -- (-1.5,-0.9) node[left] {$e^+$};
  \draw[boson] (v1) -- (v2) node[midway, above=2pt] {$Z$};
  \draw[fermion] (v2) -- ( 1.5, 0.9) node[right] {$\mu^-$};
  \draw[fermion] ( 1.5,-0.9) node[right] {$\mu^+$} -- (v2);
\end{tikzpicture}}
\end{minipage}%
\begin{minipage}[c]{0.06\textwidth}
\centering{\large$\longrightarrow$}
\end{minipage}%
\usebox{\symjsonbox}}

\end{enumerate}

The shared schema ensures that the same process, differing only in whether parameters are numerical or symbolic, feeds both calculation paths and produces directly comparable results.
This provides a \emph{single source of truth} for the physics across all toolkit stages and naturally supports future LLM-driven exploration. 

\subsection{Estimating diagrams with NDA}
\label{sec:nda}

The NDA component provides rapid order-of-magnitude estimates of decay widths and cross sections directly from the diagram specification, requiring no external software.
The approach follows the spirit of Ref.~\cite{Alwall:2014bza}, with a slightly different phase-space prescription and also adapted for cross-section estimation.
Its role is to resolve task-level parametric questions before the \EDA path is invoked, and to provide an independent scaling check on exact results after the fact.
The formalism applies equally to decays ($\Gamma \propto \Phi_n \cdot |\mathcal{M}|^2 / 2M$) and scattering ($\sigma \propto \Phi_n \cdot |\mathcal{M}|^2 / 2s$), differing only in the flux prefactor and initial-state averaging. 
For simplicity, we present the decay width form and focus on decay processes throughout this work.
The central approximation is that $\overline{|\mathcal{M}|^2}$, which generically depends on the point in phase space, is evaluated at a single characteristic energy scale $E \sim M$ and pulled out of the phase-space integral.
This renders each diagram component independent of the final-state kinematics, so that for a $1\to n$ decay of a parent with mass $M$ the width factorizes as
\begin{equation}
    \Gamma_{\text{NDA}}
    =
    \frac{1}{2M}
    \cdot \Phi_n(M)
    \cdot \overline{|\mathcal{M}|^2}_{\text{NDA}}
    \cdot C_{\text{color}},
    \label{eq:nda_master}
\end{equation}
where $C_{\text{color}}$ is the color factor.
The $n$-body phase space $\Phi_n$ uses the exact two-body result $\Phi_2 = |\vec{p}\,|/(8\pi M)$ for $n=2$; for $n \geq 3$ final-state particles assumed massless, the analytic volume~\cite{Heeck:2019kgr,Kleiss:1985gy}
\begin{equation}
    \Phi_n(M) = \frac{1}{2^{4n-5}\,\pi^{2n-3}}\,\frac{M^{2n-4}}{(n-1)!\,(n-2)!}
    \label{eq:phasespace}
\end{equation}
is used.
The squared matrix element is further factorized into a modular product of Feynman diagram components~\cite{Alwall:2014bza}, with no Lorentz contractions and no interference between diagrams:
\begin{equation}
    \overline{|\mathcal{M}|^2}_{\text{NDA}}
    \;\sim\;
    \frac{N_{\text{color}}}{N_{\text{spin}}}
    \left(\prod_{\text{vertices}} |C_i|^2\right)
    \left(\prod_{\text{internal}} |P_i|^2\right)
    \left(\prod_{\text{external}} \mathcal{P}_i\right)
    \left(\frac{1}{16\pi^2}\right)^{\!L}.
    \label{eq:nda_msq}
\end{equation}
Here $C_i$ are vertex couplings drawn from the enumerated set in \cref{tab:vertex_types} 
(shared between NDA and \EDA), $P_i = 1/(q^2 - m_{\text{prop}}^2)$ are propagator 
denominators evaluated at the characteristic scale $E \sim M$ (so that $|P_i|^2 \sim 
1/m_{\text{prop}}^4$ in the heavy regime and $|P_i|^2 \sim 1/E^4$ in the light regime), 
$\mathcal{P}_i$ are external-state polarization factors, and $L$ is the loop order.
The full prescription for propagator regimes, external-state factors, and higher-dimensional operator scaling is given in appendix~\ref{app:nda_prescription}.

Before evaluating any estimate, the NDA tools run a series of validation checks on the diagram specification, including topology consistency (vertex, propagator, and external-leg counts), graph-theory constraints ($I = (\sum n_i V_i - E)/2$), Lorentz invariance (fermions in pairs with allowed spin combinations), quantum number conservation, and kinematic threshold ($M > \sum m_i$).
These checks catch specification errors before they propagate to the calculation, providing early structured feedback to the agent on malformed inputs.

\subsection{Enumerating diagrams with \fg}
\label{sec:feyngraph}

\fg~\cite{FeynGraph} is an efficient Feynman diagram enumeration engine written in Rust with Python bindings.
It serves as a common entry point to both calculation paths, automatically enumerating all Feynman diagrams contributing to a given process (though the agent can also construct diagrams manually when the process is known in advance).
The Standard Model vertex set is available by default; BSM models can be loaded via UFO~\cite{FeynRules} model files.

Given initial- and final-state particle lists, \fg enumerates all topologically distinct diagrams consistent with the model vertices.
The \adname wrapper around \fg then converts each enumerated diagram to the shared \texttt{Diagram} dataclass of \cref{sec:diagram_spec} via automatic vertex-type inference, mapping SM particle content to the vertex vocabulary of \cref{tab:vertex_types} without requiring the agent to know the Lorentz structure in advance (e.g., a $W$ boson coupled to a fermion pair is assigned \texttt{"left-handed"}, a $Z$ boson \texttt{"vector-axial"}).

The converted diagrams are ranked by physics importance using a scoring function implemented in \adname.
The ranking is dominated by the number of heavy internal propagators, since each propagator with mass $M_{\text{prop}} \gg E$ contributes a suppression factor $(E/M_{\text{prop}})^4$ to $|\mathcal{M}|^2$ (from squaring the propagator denominator at the characteristic scale).
The ranking score
\begin{equation}
    \text{score} = 10^{\,3 - 2n_{\text{loops}} - 3n_{\text{heavy}}} \cdot \prod_i g_i^2
    \label{eq:ranking}
\end{equation}
penalizes loop order ($100^{n_{\text{loops}}}$), weights by vertex coupling strengths ($\prod_i g_i^2$), and applies a proxy suppression per heavy propagator ($10^{-3n_{\text{heavy}}}$).
The proxy factor $10^{-3}$ is chosen conservatively to ensure that diagrams with fewer heavy propagators always rank above those with more, regardless of coupling variations.

The \adname enumeration tool organizes the output into classes grouped by heavy-propagator count, with each class directory containing individual diagram JSON files, visualizations, and a representative diagram suitable for passing directly to the NDA or \EDA tools.
A per-class summary table recording diagram counts and coupling structures is saved alongside the enumeration, providing the structured metadata that downstream tools and plotting scripts consume.

\subsection{Computing diagrams with EDA}
\label{sec:diagrammatica}

The \EDA path generates complete, self-contained Mathematica/\fc scripts from symbolic diagram specifications, executes them via \texttt{wolframscript}, and returns parsed structured results.
The code generator maps the abstract diagram description to concrete \fc function calls covering the procedure outlined in \cref{sec:token_entropy}, handling three topology classes: direct $1\to 2$ decays, $1\to 2$ decays with one internal propagator, and $2\to 2$ scattering with $s$-, $t$-, or $u$-channel exchange.

Each call computes a single diagram for a specific set of particles, a single vertex type, and a single coupling symbol per chiral projection.
When different final-state flavors do not interfere (e.g., $Z \to e^+e^-$ vs.\ $Z \to \mu^+\mu^-$), inclusive widths are assembled by the agent through multiple tool calls, keeping each individual computation simple and auditable.
A notable limitation of this single-coupling-per-vertex design is that it does not capture the flavor structure arising from non-diagonal coupling matrices.
For example, a scalar $\phi$ with a general Yukawa interaction $c_{ij} \phi \bar{\psi}_i \psi_j + {\rm h.c.}$ induces $\psi_j \to \psi_i \phi$ through two interfering amplitudes proportional to $c_{ij}$ and $c_{ji}^*$ respectively, producing a width $\Gamma \propto (|c_{ij}|^2 + |c_{ji}|^2)(\ldots) + \mathrm{Re}(c_{ij}c_{ji})(\ldots)$ where the cross term between distinct matrix entries cannot be obtained from a single coupling symbol.
Extending the vertex specification to accept flavor-indexed coupling matrices, and the code generator to coherently sum the resulting amplitudes before squaring, would enable the toolkit to derive such expressions directly. 
We leave this for future work.
The vertex type field of the diagram specification maps to a concrete Lorentz structure inserted into the \fc amplitude (\cref{tab:vertex_types}).
The design is model-independent, with vertex types specified by their Lorentz structure (e.g., \texttt{"vector-axial"} for $\gamma^\mu(g_V - g_A\gamma_5)$) rather than by SM-specific identifiers, so the same toolkit applies without modification to BSM theories sharing the same Lorentz structures.

\begin{table}[t!]
\centering
\begin{tabular}{lllll}
\toprule
\textbf{Family} & \textbf{Type} & \textbf{Lorentz structure} & $|\mathcal{M}|^2$ & \textbf{\fc code} \\
\midrule
\multirow{4}{*}{SFF}
 & \texttt{scalar}       & $\mathbb{1}$              & $y^2 M^2$ & \texttt{I g} \\
 & \texttt{pseudoscalar} & $\gamma_5$                & $y^2 M^2$ & \texttt{I g GA[5]} \\
 & \texttt{chiral}       & $g_L P_L + g_R P_R$       & $y^2 M^2$ & \texttt{I(gL GA[7]+gR GA[6])} \\
 & \texttt{scalar-va}    & $g_S + g_P\gamma_5$       & $y^2 M^2$ & \texttt{I(gS+gP GA[5])} \\
\midrule
\multirow{8}{*}{VFF}
 & \texttt{vector}       & $\gamma^\mu$                    & $g^2 M^2$ & \texttt{I g GAD[mu]} \\
 & \texttt{axial-vector} & $\gamma^\mu\gamma_5$            & $g^2 M^2$ & \texttt{I g GAD[mu].GA[5]} \\
 & \texttt{left-handed}  & $\gamma^\mu P_L$                & $g^2 M^2$ & \texttt{I g GAD[mu].GA[7]} \\
 & \texttt{right-handed} & $\gamma^\mu P_R$                & $g^2 M^2$ & \texttt{I g GAD[mu].GA[6]} \\
 & \texttt{vector-axial} & $\gamma^\mu(g_V - g_A\gamma_5)$ & $g^2 M^2$ & \texttt{I GAD[mu].(gV-gA GA[5])} \\
 & \texttt{chiral}       & $\gamma^\mu(g_L P_L + g_R P_R)$ & $g^2 M^2$ & \texttt{I GAD[mu].(gL GA[7]+gR GA[6])} \\
 & \texttt{tensor}       & $\sigma^{\mu\nu}k_\nu$          & $g^2 M^4$ & \texttt{I g DiracSigma[...]} \\
 & \texttt{tensor-chiral} & $(g_L P_L\!+\!g_R P_R)\sigma^{\mu\nu}k_\nu$ & $g^2 M^4$ & \texttt{I(gL GA[7]+gR GA[6]).DiracSigma[...]} \\
\midrule
\multirow{4}{*}{Bosonic}
 & SSS         & $ig$               & $\lambda^2$   & \texttt{I g} \\
 & SSV         & $ig(p_1-p_2)^\mu$  & $g^2 M^2$     & \texttt{I g (p1-p2)$\cdot\varepsilon$} \\
 & SVV (dim-4) & $ig\,g^{\mu\nu}$   & $g^2 M^2$     & \texttt{I g MT[mu,nu]} \\
 & VVV         & triple gauge       & $g^2 M^2$     & \texttt{g f$^{abc}$[(p1-p2)g+cyc]} \\
\midrule
\multirow{2}{*}{Dim-5}
 & $\phi FF$         & $ig(k_1\!\cdot\!k_2\,g^{\mu\nu} - k_1^\nu k_2^\mu)$ & $g^2 M^4$ & field-strength \\
 & $\phi F\tilde{F}$ & $g\,\epsilon^{\mu\nu\rho\sigma}k_{1\rho}k_{2\sigma}$ & $g^2 M^4$ & dual field-strength \\
\midrule
\multirow{2}{*}{EFT}
 & \texttt{dim5-weinberg} & $(1/\Lambda)\,LLHH$        & $M^2/\Lambda^2$ & --- \\
 & \texttt{dim6-4fermion} & $(1/\Lambda^2)(\bar\psi\psi)^2$ & $M^4/\Lambda^4$ & --- \\
\bottomrule
\end{tabular}
\caption{Unified vertex type system shared by the NDA and \EDA paths.
SFF = scalar--fermion--fermion; VFF = vector--fermion--fermion.
The $|\mathcal{M}|^2$ column gives the NDA scaling; the \fc column gives the exact Lorentz structure used in code generation.
Aliases (e.g., \texttt{yukawa} for \texttt{scalar}, \texttt{va} for \texttt{vector-axial}, \texttt{dipole} for \texttt{tensor}) are accepted by both paths.
EFT operators are NDA-only (no code generation).}
\label{tab:vertex_types}
\end{table}

Each generated script emits structured output through labeled markers that the tool runner parses automatically: symbolic expressions in Mathematica \texttt{InputForm}, \LaTeX-formatted formulas via \texttt{TeXForm}, and numerical evaluations.
These results are persisted in JSON \emph{sidecar} files (companion files saved alongside the generated \texttt{.wl} scripts with a \texttt{\_results.json} suffix), enabling downstream tools to consume results by reference rather than requiring the agent to copy expressions manually.
A typical \EDA workflow chains 3--5 tool calls. 
After the agent specifies a \texttt{Diagram} (either by hand or via FeynGraph), \textbf{\textsf{ComputeSymbolicAmplitude}} generates the script, \textbf{\textsf{RunWolframScript}} executes it, optional \textbf{\textsf{SimplifyResult}} calls apply limits or substitutions, and \textbf{\textsf{ConvertToPython}} exports the final formula as a callable Python function.

\subsection{Theory knowledge base}
\label{sec:skills_graph}

\adname includes a navigable QFT knowledge base, the \emph{skills graph}, exposed through the \textbf{\textsf{LookupTheory}} tool.
This is the $\Delta_C$-reducing component of the toolkit identified in \cref{sec:reliability}: rather than flooding the agent's context with reference material, the knowledge base exposes exactly the document the agent needs at each decision point.

The graph comprises 25 interlinked documents organized into seven categories (Feynman rules, trace identities, spin sum formulas, phase-space conventions, \fc reference, procedures, and worked examples), each pairing theoretical content with working \fc code.
The \textbf{\textsf{LookupTheory}} tool exposes actions for retrieving individual documents by key, searching across all nodes, navigating links between related topics, and listing nodes by category.
The complete knowledge base and source code are available in the public code release (\cref{sec:data_availability}).

\section{Benchmarks}
The two benchmark tasks defined in this work serve complementary roles in validating the architecture of \cref{sec:architecture}.
Task~1 (\cref{sec:catalog}) exercises the \EDA path on a systematic sweep of all supported vertex families, testing the shared diagram specification, code generator, and multi-fidelity cross-checks.
Task~2 (\cref{sec:muon}) exercises the NDA path and \fg enumeration engine on a problem with strong combinatorial growth, stressing diagram ranking and approximate rate estimation across large diagram classes.
Both tasks are presented through curated, annotated agent transcripts from single sessions. 
Extended agent transcripts can be found in appendices~\ref{app:task1_transcript} and~\ref{app:task2_transcript}, while the full system prompts and complete, unedited conversation logs are available in the accompanying data release (see \cref{sec:data_availability}).

Throughout this work, we use Claude Opus~4.6~\cite{Claude} as the LLM agent via Claude Code, Anthropic's CLI for agentic development.
Like \pkgname~\cite{Menzo:2025cim}, the \adname toolkit is implemented in the Orchestral AI framework~\cite{roman2026orchestralaiframeworkagent}, which supports all major LLM providers and allows for all tools to be exposed via the Model Context Protocol (\MCP)~\cite{MCP}.
Since \MCP has been adopted as the standard tool-integration protocol across the major agentic frameworks (LangChain/LangGraph, OpenAI Agents SDK, CrewAI, Microsoft Agent Framework, and PydanticAI), the same tools are immediately consumable by any MCP-compatible agent, independent of the underlying model provider.

Each task is run in a fresh sandbox with a task-specific subset of \pkgname tools pre-authorized and a structured system prompt documenting the tool interfaces and available workflows.
The system prompts do not contain task-specific instructions; the agent must plan its own approach from the combination of the task prompt and the tool documentation.

\subsection{Task 1: Exhaustive \texorpdfstring{$1\to 2$}{1 -> 2} decay rate catalog}
\label{sec:catalog}

This section validates the \EDA path on a systematic benchmark, the complete set of tree-level, single-vertex $1\to 2$ decays supported by the toolkit.
The benchmark verifies that the shared diagram specification, vertex type system, and symbolic code generator cover the intended phenomenological space in a uniform way, and demonstrates that the agent, given only a task description and tool access, can autonomously plan, execute, and cross-check a multi-step symbolic computation spanning every supported vertex family.

\subsubsection{The toolbox}
The agent was given access to the following \pkgname tools, in addition to the native Claude Code tools (\textbf{\textsf{Read}}, \textbf{\textsf{Write}}, \textbf{\textsf{Edit}}, \textbf{\textsf{Bash}}, \textbf{\textsf{Glob}}, \textbf{\textsf{ToolSearch}}):
\begin{toolbox}{Toolbox (Task 1)}
\begin{itemize}[leftmargin=1.5em, itemsep=1pt, topsep=2pt]
    \item \textbf{\textsf{ComputeSymbolicAmplitude}} --- generate \fc WolframScript from a diagram specification
    \item \textbf{\textsf{RunWolframScript}} / \textbf{\textsf{RunWolframScriptBatch}} --- execute scripts via \fc
    \item \textbf{\textsf{SimplifyResult}} / \textbf{\textsf{SimplifyResultBatch}} --- apply substitutions and limits to symbolic results
    \item \textbf{\textsf{ConvertToPython}} --- export symbolic formulas as Python callables
    \item \textbf{\textsf{EstimateDecayWidthNDA}} --- NDA cross-check estimates
    \item \textbf{\textsf{PDGDatabase}} --- particle properties and coupling data
\end{itemize}
\end{toolbox}

\subsubsection{The user prompt}
\vspace{0.1in}

\begin{agentuser}
Enumerate and compute the symbolic partial decay width $\Gamma(A \to BC)$ for every tree-level, Lorentz-invariant, single-vertex $1 \to 2$ process in four spacetime dimensions. Systematically sweep over parent spins $\{0, \frac{1}{2}, 1\}$ and all kinematically allowed daughter spin assignments. Retain complex couplings and distinct daughter masses throughout; equal-mass, massless, and real-coupling limits are taken as special cases. For each vertex type that admits both a vector--axial $(g_V, g_A)$ or scalar--pseudoscalar $(g_S, g_P)$ parameterization and a chiral $(g_L, g_R)$ parameterization, compute both in parallel and verify algebraic consistency under the basis change.
Validate a representative subset of the computed formulas against known Standard Model partial widths, using known masses and couplings.
\vspace{0.1in}

\textbf{\textit{Deliverable}}: a complete reference table of independent decay-width formulas in both coupling bases where applicable, together with useful limits, chiral cross-checks, NDA cross-checks, and Standard Model validation. Include figures that make the physical content of the catalog accessible and intuitive --- the key patterns and threshold structure should be assessable from the figures alone. Finally, provide a summary of phenomenological insights from the analysis.
\end{agentuser}

\subsubsection{Workflow}
\label{sec:task1_workflow}

\begin{figure}[t!]
\centering
\includegraphics[width=0.8\textwidth]{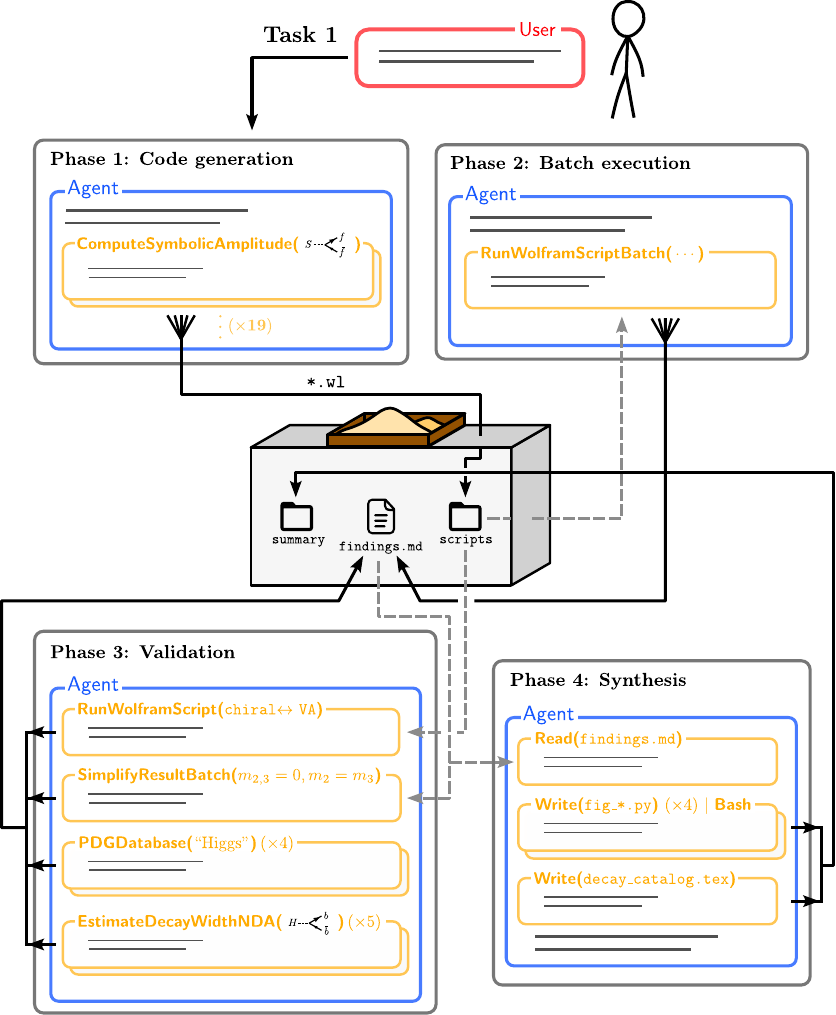}
\caption{Schematic summary flowchart of a common four-phase solution mode for the task 1 benchmark. A full transcript can be seen in appendix~\ref{app:task1_transcript}. 
}
\label{fig:flowchart_task1}
\end{figure}

Given only this prompt and the tool documentation, the agent autonomously planned and ``one-shotted'' \textit{i.e.~}executed the entire catalog in a single session with no further human input.
The session comprised 66 tool calls over 29 minutes of LLM inference time, generating ${\sim}\,67\text{k}$ output tokens against ${\sim}\,9\text{M}$ cumulative input tokens.
The workflow, illustrated in \cref{fig:flowchart_task1}, proceeds through four phases.

\begin{enumerate}[leftmargin=2em]

\item \textbf{Enumeration and code generation:}
The agent enumerated the full process space: all kinematically allowed single-vertex $1 \to 2$ decays across parent spins $\{0, \frac{1}{2}, 1\}$.
It identified 19 independent processes spanning 6 vertex families (SFF, VFF, SSS, SSV, SVV including dim-5 $\phi FF$ and $\phi F\tilde{F}$ operators, and VVV), and organized them into batches by parent spin.
For each vertex family admitting dual parameterizations---scalar--pseudoscalar versus chiral for SFF, and vector--axial versus chiral for VFF---the agent computed both bases as separate \textbf{\textsf{ComputeSymbolicAmplitude}} calls, yielding 20 code-generation calls producing 19 distinct \fc scripts.
Each call specifies the process through the structured diagram JSON of \cref{sec:diagram_spec}: spins, particle labels, and a vertex type string that selects the Lorentz structure from \cref{tab:vertex_types}.
The code generator handles all downstream algebra---Dirac traces, spin sums, polarization sums, phase space, and kinematic simplification---producing a self-contained WolframScript with structured output markers.

\item \textbf{Batch execution:}
All 19 scripts were dispatched in a single \textbf{\textsf{RunWolframScriptBatch}} call with 5-way concurrency.
One script (the $V \to f\bar{f}$ tensor vertex) timed out on the first attempt; the agent diagnosed the issue, regenerated the script with a simplified diagram specification via a fresh \textbf{\textsf{ComputeSymbolicAmplitude}} call, and retried successfully---an example of the self-correction loop enabled by the tool-constrained architecture, where failures produce structured error messages that the agent can act on rather than silent wrong answers.

\begin{figure}[t!]
\centering
\includegraphics[width=\textwidth]{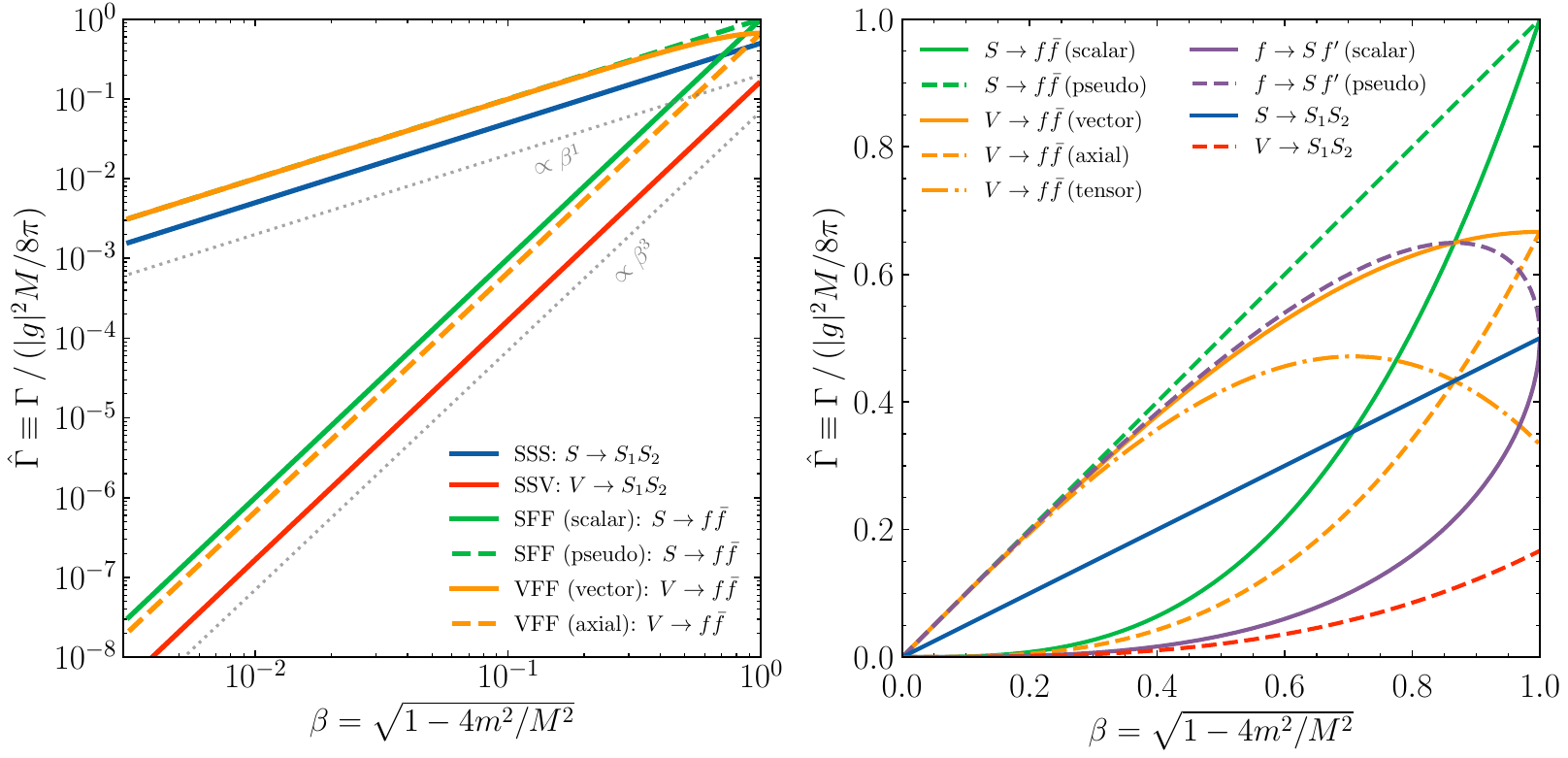}\\[10pt]
\includegraphics[width=0.6\textwidth]{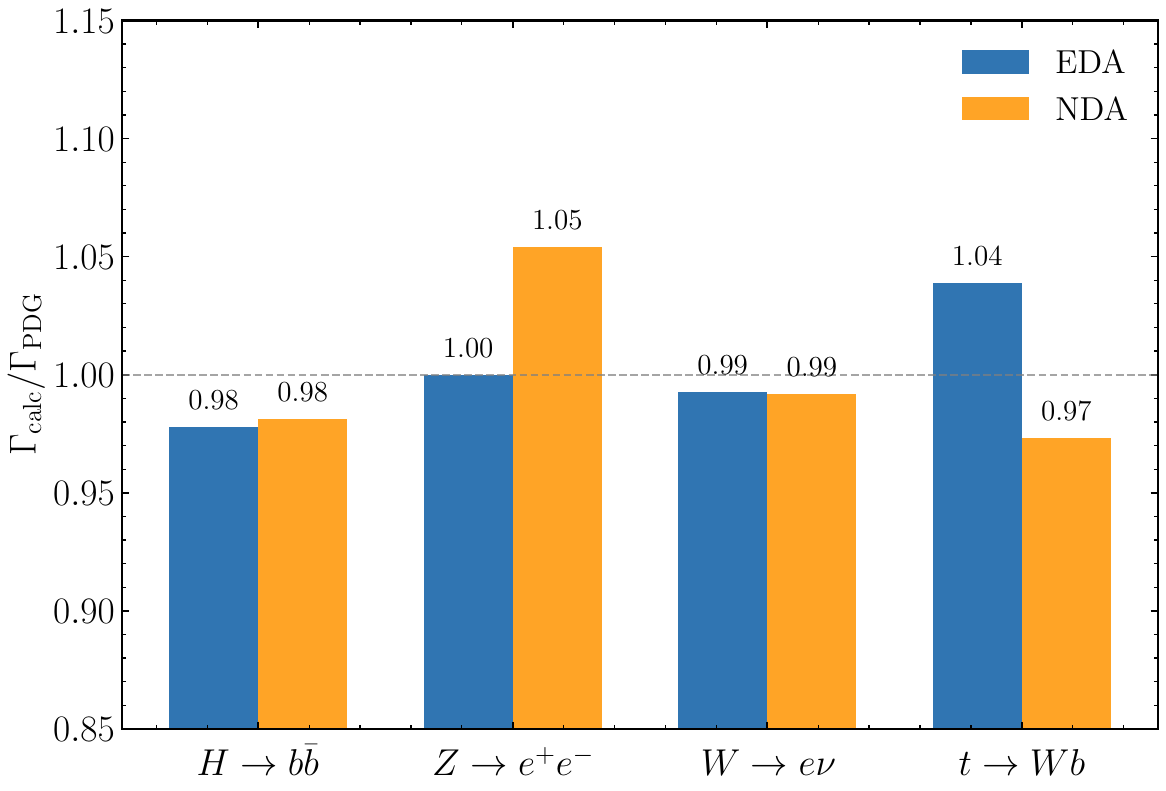}
\caption{Threshold structure and Standard Model validation of the decay-width catalog.
\textbf{Top left:} Log-log plot of the normalized widths $\hat{\Gamma} \equiv \Gamma/(|g|^2 M/8\pi)$ revealing the power-law onset $\hat{\Gamma} \propto \beta^n$ near threshold, cleanly separating S-wave ($\beta^1$) from P-wave ($\beta^3$) channels.
\textbf{Top right:} Full kinematic range on linear scale, showing the crossover between scalar and pseudoscalar SFF channels and the inclusion of spin-$\frac{1}{2}$ parent decays.
\textbf{Bottom:} Ratio of computed to PDG partial widths for four benchmark decays; tree-level exact formulas (blue) agree within 2--4\%, NDA improved estimates (orange) within $\sim\!5\%$.
\textit{Figures generated by the agent with minor cosmetic adjustments.}}
\label{fig:catalog_summary}
\end{figure}

\item \textbf{Validation:}
With all 19 formulas in hand, the agent entered a systematic validation phase.
It ran six chiral $\leftrightarrow$ VA cross-checks via \textbf{\textsf{RunWolframScript}}, confirming all algebraic differences to be exactly zero in Mathematica.
It computed 17 special-case limits (equal-mass, massless) via two \textbf{\textsf{SimplifyResultBatch}} calls, queried \textbf{\textsf{PDGDatabase}} for SM masses and couplings ($m_H$, $m_W$, $m_Z$, $m_t$, $\sin^2\theta_W$, $G_F$, and Yukawa couplings), and ran five parallel \textbf{\textsf{EstimateDecayWidthNDA}} calls to cross-check the exact results against order-of-magnitude dimensional analysis.
This multi-fidelity cross-check---exact symbolic formulas validated against both experimental data and an independent approximate method---is the concrete realization of the design principle identified in \cref{sec:architecture}: the same diagram specification feeds both paths, and agreement in scaling confirms internal consistency.

\item \textbf{Synthesis:}
The agent wrote four standalone Python plotting scripts (executed to generate publication figures) and compiled a 12-page \LaTeX reference document containing all 19 formulas in both coupling bases, the cross-check and validation tables, and a phenomenological analysis of the catalog.
The complete reference document is reproduced in appendix \ref{app:catalog}.

\end{enumerate}

\subsubsection{Results}
\label{sec:catalog_results}

\Cref{tab:catalog} summarizes the 19 independent formulas computed by the agent, listing the process, vertex type, threshold power law, and massless limit for each. 
Full formulas with distinct daughter masses are collected in appendix~\ref{app:catalog}.
For dimensionless couplings the universal pattern is $\Gamma \propto |g|^2 M/(8\pi)$, modulated by spin averaging and angular momentum barriers; dimension-1 couplings (SSS) give $\Gamma \propto |g|^2/M$; dimension-$(-1)$ operators (tensor, $\phi FF$) give $\Gamma \propto |g|^2 M^3$.
Three channels ($S \to V_1V_2$ via SVV, $V \to SV'$, $V \to V_1V_2$) diverge in the massless-vector limit, reflecting the breakdown of perturbative unitarity from longitudinal polarization growth. 

Interestingly for the spin-$\frac{1}{2}$ parent family, the agent consistently specified the fermion-sector daughter as an antifermion, choosing to compute $f_1 \to \bar{f}_2\,X$ rather than the fermion-number-conserving $f_1 \to f_2\,X$ that a traditionally trained particle physicist might write down first.
At the amplitude level, these correspond to two distinct spinor contractions that are both Lorentz invariant:
\begin{equation}
    \bar{u}(p_{f_2})\,\Gamma\, u(p_{f_1})
    \;\longleftrightarrow\;
    f_1 \to f_2\, X,\quad \Delta F = 0,
    \qquad
    \bar{v}(p_{f_2})\,\Gamma\, u(p_{f_1})
    \;\longleftrightarrow\;
    f_1 \to \bar{f}_2\, X,\quad \Delta F = 2,
\end{equation}
where the $v$ spinor satisfies the Dirac equation with opposite-sign eigenvalue and is related to $u$ by charge conjugation, $v = C\bar{u}^T$.
In the Lagrangian, the $\bar{v}\Gamma u$ amplitude corresponds to a Majorana-type coupling $\psi_{f_2}^T C\,\Gamma\,\psi_{f_1}\,X + \text{h.c.}$, which violates fermion number by two units.
Both structures are legitimate entries in a complete Lorentz-invariant enumeration.
The agent's preference for $\bar{v}\Gamma u$ reflects that it imposes no prior on fermion number conservation, enumerating all Lorentz-invariant amplitudes without privileging the symmetries of the Standard Model.
Ultimately, this reflects an interesting and revealing misalignment between the agent's action distribution and the implicit prior a trained particle physicist would bring to the same task.
It also reveals that a subtle modification to the prompt, such as asking the agent to assume compatibility with a global $U(1)_F$, would have meaningfully shifted the agent's action distribution toward what we might call ``physics alignment'' \textit{i.e.}~the degree to which an agent's action distribution reflects the conventions, priors, and symmetry assumptions a domain expert would implicitly impose.

Additionally, when the agent subsequently attempted to validate $t \to Wb$ using the $f_1 \to \bar{f}_2 V$ topology, the pre-computation checks in \textbf{\textsf{EstimateDecayWidthNDA}} flagged the assignment $t \to \bar{b}\,W^+$ as a charge-conservation violation.
The agent was prompted to re-specify the process as $t \to b\,W^+$ before proceeding.
This episode underscores the necessity of structured validation checks with informative error messages, as a physically incorrect assignment that the agent would have otherwise propagated silently was caught and corrected without manual intervention.

\begin{agentresponse}
\centering
\small
\begin{tabular}{lllcl}
\toprule
\textbf{Process} & \textbf{Vertex} & \textbf{Threshold} & \textbf{Massless limit} & \textbf{Couplings} \\
\midrule
\multicolumn{5}{l}{\textit{Spin-0 parent}} \\[2pt]
$S\to f\bar{f}$ & scalar ($g_S$) & $\beta^3$ & $|g_S|^2 m_S/(8\pi)$ & 2 \\
$S\to f\bar{f}$ & pseudoscalar ($g_P$) & $\beta^1$ & $|g_P|^2 m_S/(8\pi)$ & 2 \\
$S\to f\bar{f}$ & scalar-VA ($g_S,g_P$) & mixed & $(|g_S|^2\!+\!|g_P|^2)\, m_S/(8\pi)$ & 4 \\
$S\to f\bar{f}$ & chiral ($g_L,g_R$) & mixed & $(|g_L|^2\!+\!|g_R|^2)\, m_S/(16\pi)$ & 4 \\
$S\to S_1S_2$ & SSS & $\beta^1$ & $|g|^2/(16\pi m_S)$ & 2 \\
$S\to V_1V_2$ & SVV & $\beta^1$ & $\infty$ (unitarity) & 2 \\
$S\to V_1V_2$ & $\phi FF$ & $\beta^1$ & $|g|^2 m_S^5/(8\pi)$ & 2 \\
$S\to V_1V_2$ & $\phi F\tilde{F}$ & $\beta^3$ & $|g|^2 m_S^5/(8\pi)$ & 2 \\[4pt]
\multicolumn{5}{l}{\textit{Spin-$\frac{1}{2}$ parent}} \\[2pt]
$f\to S\bar{f}'$ & scalar-VA & mixed & $(|g_S|^2\!+\!|g_P|^2)\, m_f/(16\pi)$ & 4 \\
$f\to S\bar{f}'$ & chiral & mixed & $(|g_L|^2\!+\!|g_R|^2)\, m_f/(32\pi)$ & 4 \\
$f\to V\bar{f}'$ & VA ($g_V,g_A$) & $\beta^1$ & $(|g_V|^2\!+\!|g_A|^2)\,m_f/(8\pi)$ & 4 \\
$f\to V\bar{f}'$ & chiral ($g_L,g_R$) & $\beta^1$ & $(|g_L|^2\!+\!|g_R|^2)\,m_f/(16\pi)$ & 4 \\
$f\to V\bar{f}'$ & tensor ($g$) & $\beta^1$ & $|g|^2 m_f^3/(16\pi)$ & 2 \\
$f\to V\bar{f}'$ & tensor-chiral & $\beta^1$ & \mbox{see \cref{app:catalog}} & 4 \\[4pt]
\multicolumn{5}{l}{\textit{Spin-1 parent}} \\[2pt]
$V\to f\bar{f}$ & VA ($g_V,g_A$) & $\beta^1$ & $(|g_V|^2\!+\!|g_A|^2)\, m_V/(12\pi)$ & 4 \\
$V\to f\bar{f}$ & chiral ($g_L,g_R$) & $\beta^1$ & $(|g_L|^2\!+\!|g_R|^2)\, m_V/(24\pi)$ & 4 \\
$V\to f\bar{f}$ & tensor ($g$) & $\beta^1$ & $|g|^2 m_V^3/(24\pi)$ & 2 \\
$V\to f\bar{f}$ & tensor-chiral & $\beta^1$ & \mbox{see \cref{app:catalog}} & 4 \\
$V\to S_1S_2$ & SSV & $\beta^3$ & $|g|^2 m_V/(48\pi)$ & 2 \\
$V\to SV'$ & SVV & $\beta^1$ & $\infty$ (unitarity) & 2 \\
$V\to V_1V_2$ & VVV & $\beta^1$ & $\infty$ (unitarity) & 2 \\
\bottomrule
\end{tabular}
\captionof{table}{Complete catalog of $1\to 2$ tree-level decay widths computed by the agent, organized by parent spin.
$\beta = \sqrt{1-4m^2/M^2}$ for equal-mass daughters.
``Couplings'' counts independent real parameters per complex coupling.}
\label{tab:catalog}
\end{agentresponse}

The SM validation (\cref{tab:sm_validation}) also demonstrates that the generic formulas, evaluated at PDG masses and couplings, reproduce known partial widths to within the few-percent accuracy expected at tree level.
The $H\to b\bar{b}$ result uses the running mass $m_b(m_H) \approx 2.79$~GeV with colour factor $N_c = 3$; the $t\to Wb$ tree-level overestimate of $\sim\!4\%$ is consistent with the known $\sim\!9\%$ NLO QCD correction.
The NDA column confirms internal consistency between the two calculation paths: the same diagram specification yields compatible results from independent methods, with the largest deviation ($\sim\!6\%$ for $Z \to e^+e^-$) arising from the $g_V/g_A$ asymmetry that NDA does not resolve.

\begin{agentresponse}
\centering
\begin{tabular}{llccc}
\toprule
Process & Formula & Tree-level & PDG & Ratio \\
\midrule
$H\to b\bar{b}$ & Eq.~\eqref{eq:S-ff-va}, $g_S=m_b/v$, $N_c=3$
  & $1.91$ MeV & $1.96$ MeV & $0.98$ \\
$Z\to e^+e^-$ & Eq.~\eqref{eq:V-ff-massless}, $g_V^e, g_A^e$
  & $83.9$ MeV & $83.9$ MeV & $1.00$ \\
$W\to e\nu$ & Eq.~\eqref{eq:V-ff-ch}, $g_L=g/\sqrt{2}$, $g_R=0$
  & $227.2$ MeV & $228.9$ MeV & $0.99$ \\
$t\to Wb$ & Eq.~\eqref{eq:f-Vf-ch}, $g_L=g/\sqrt{2}$, $g_R=0$
  & $1.479$ GeV & $1.424$ GeV & $1.04$ \\
\bottomrule
\end{tabular}
\captionof{table}{Standard Model validation.  Tree-level symbolic widths evaluated at PDG masses and couplings, compared with experimental partial widths.  All four agree within $\sim\!4\%$; the residual discrepancy reflects higher-order corrections absent at tree level.}
\label{tab:sm_validation}
\end{agentresponse}

\subsubsection{Phenomenological analysis}
\label{sec:catalog_pheno}

Beyond computing the formulas, the agent identified several physical patterns that emerge from surveying the full catalog.
The threshold behavior (\cref{fig:catalog_summary}) classifies all vertices by their near-threshold power law, and the agent's analysis connects this classification to the underlying angular momentum structure:

\begin{agentresponse}
\textit{\textbf{1.\quad Threshold classification by partial wave}}
\smallskip

The near-threshold behavior $\Gamma \propto \beta^{2\ell+1}$ encodes the orbital angular momentum $\ell$ of the decay products. From our catalog:

\textbf{S-wave} ($\beta^1$): SSS, pseudoscalar SFF ($S\to f\bar{f}$), all VFF gauge vertices ($V\to f\bar{f}$, $f\to V\bar{f}'$), SVV and $\phi FF$ operators, VVV.
These dominate near threshold and are the first channels to ``turn on'' above the kinematic limit.

\textbf{P-wave} ($\beta^3$): scalar SFF ($S\to f\bar{f}$), SSV ($V\to SS$), $\phi F\tilde{F}$.
These are suppressed near threshold by centrifugal barrier effects.
The P-wave nature of $V\to S_1 S_2$ in the SSV vertex comes from the derivative coupling $g\,\varepsilon\cdot(p_1-p_2)$.

The scalar-vs-pseudoscalar threshold dichotomy in SFF is a direct consequence of parity: a $0^+$ scalar decaying to a fermion--antifermion pair requires $\ell=1$ (P-wave, $\beta^3$), while a $0^-$ pseudoscalar allows $\ell=0$ (S-wave, $\beta^1$).
\end{agentresponse}

\noindent The agent derived a CP discriminant directly from the equal-mass limit of the SFF formulas:

\begin{agentresponse}
\textit{\textbf{3.\quad Scalar vs.\ pseudoscalar discrimination}}
\smallskip

The ratio $\Gamma_{g_S}/\Gamma_{g_P}$ at fixed $|g|^2$ provides a clean discriminant of the CP nature of a scalar coupling:
$$\frac{\Gamma_{g_S}}{\Gamma_{g_P}} = \frac{m_S^2-4m_f^2}{m_S^2} = \beta^2\,.$$
Near threshold ($\beta\to 0$), the pseudoscalar coupling dominates by a factor $1/\beta^2$. Far above threshold ($\beta\to 1$), the scalar and pseudoscalar rates coincide---the chirality flip becomes kinematically inexpensive. This ratio is observable through angular distributions and threshold scans.
\end{agentresponse}

\noindent The agent also identified a longitudinal polarization enhancement in channels with massive vector daughters:

\begin{agentresponse}
\textit{\textbf{2.\quad Longitudinal mode enhancement}}
\smallskip

Channels with massive vector bosons in the final state exhibit growth $\propto M^{2n}/m_V^{2n}$ for $M \gg m_V$, driven by the longitudinal polarization vector $\varepsilon_L^\mu \sim k^\mu/m_V$. Specifically:
$S\to V_1V_2$ (SVV) grows as $m_S^4/(4m_{V_1}^2 m_{V_2}^2)$, corresponding to $\Gamma \propto m_S^5/m_V^4$ at large $m_S/m_V$; $V\to V_1V_2$ (VVV) shows even steeper growth, $\Gamma \propto m_V^5/m_{V_1}^4$.
This growth signals the breakdown of perturbative unitarity: a renormalizable coupling $g$ of dimension [mass] produces a width growing faster than $M$, eventually violating the bound $\Gamma < M$. In the Standard Model, this is cured by the Higgs mechanism, which relates $g \propto m_V$, so that $\Gamma/M$ remains finite.
\end{agentresponse}

\noindent Finally, a universal feature noted across all VFF vertices is the interference term $\pm 6\, m_f m_{\bar{f}} m_V^2$ that distinguishes vector from axial couplings:

\begin{agentresponse}
\textit{\textbf{4.\quad The $6m_f m_{\bar{f}} m_V^2$ interference term}}
\smallskip

A universal feature of VFF vertices is the interference term $\pm 6\, m_f m_{\bar{f}} m_V^2$ that distinguishes vector from axial (or $g_L$ from $g_R$) couplings. This term vanishes for massless fermions, making left- and right-handed rates identical (as required by chirality conservation); is maximized when both fermion masses are comparable to $m_V$; and carries opposite signs for vector vs.\ axial couplings, enabling their separation through mass-dependent rate measurements.
\end{agentresponse}

\noindent These observations---the partial-wave classification, the CP discriminant, the unitarity-violating growth, and the $V$-$A$ interference structure---are not individually novel, but the fact that they were identified autonomously by the agent from a systematic survey of the computed formulas illustrates the kind of pattern recognition and utility that becomes possible when symbolic computation is reliable enough to trust.

\subsection{Task 2: Muon decay multiplicity sensitivity study}
\label{sec:muon}

This section validates the enumeration and NDA layers on a benchmark with strong combinatorial growth.
While Task~1 exercised the \EDA path on a systematic sweep of simple topologies, Task~2 stresses automatic diagram generation, ranking, and approximate rate estimation across large diagram classes.
It also demonstrates the multi-fidelity cross-check between NDA and MadGraph on identical diagram specifications.

\subsubsection{The toolbox}
The agent was given access to a different \pkgname tool set focused on the NDA path, alongside the native Claude Code tools (\textbf{\textsf{Read}}, \textbf{\textsf{Write}}, \textbf{\textsf{Bash}}, \textbf{\textsf{Glob}}, \textbf{\textsf{Agent}}, \textbf{\textsf{WebSearch}}):
\begin{toolbox}{Toolbox (Task 2)}
\begin{itemize}[leftmargin=1.5em, itemsep=1pt, topsep=2pt]
    \item \textbf{\textsf{PDGSearch}} --- particle properties and experimental data
    \item \textbf{\textsf{EnumerateDiagrams}} --- automatic diagram enumeration and ranking
    \item \textbf{\textsf{EstimateBranchingRatioNDA}} / \textbf{\textsf{EstimatePhaseSpace}} --- NDA estimates
    \item \textbf{\textsf{MadGraphFromRunCard}} --- exact tree-level widths via MadGraph~5
\end{itemize}
\end{toolbox}

\subsubsection{The user prompt}
\begin{agentuser}
Determine the largest number of $e^+e^-$ pairs $n$ for which the SM decay $\mu^+ \to \bar{\nu}_\mu \nu_e + n(e^+e^-) + e^+$ remains observable at current or planned muon experiments.  Begin by establishing the experimental landscape: existing measurements, planned experiments (Mu3e, etc.), and their sensitivity reaches.  For each multiplicity, enumerate the contributing tree-level diagrams and estimate the branching ratio using NDA.  Continue increasing $n$ until the rate falls below experimental sensitivity.
\vspace{0.1in}

Cross-check the dominant diagram class at the lowest and next-to-lowest non-trivial multiplicity against an exact MadGraph calculation, using coupling order restrictions to isolate the relevant diagrams.  Provide a brief analytical estimate of the per-pair suppression factor and compare it with the observed numerical scaling.
\vspace{0.1in}

\textbf{\textit{Deliverable}}: a comprehensive table of diagram counts, NDA widths, scalings, and branching ratios for each final-state multiplicity, identifying the experimental frontier.  Include ``diagram metadata'' figures that make the enumeration and NDA results intuitive.  Additionally create a summary figure illustrating how the total branching ratio scales with $n$ relative to experimental sensitivity thresholds.  Conclude with a discussion of the physics driving the multiplicity scaling and its implications for current and planned experiments.
\end{agentuser}

\subsubsection{Workflow}
\label{sec:task2_workflow}

The agent also completed Task~2 in one-shot from a single user prompt and no further human input, dispatching 63 tool calls (plus 39 from a research subagent) over 38 minutes of LLM inference time, generating ${\sim}\,54\text{k}$ output tokens against ${\sim}\,7.9\text{M}$ cumulative input tokens.
The workflow proceeds through four phases illustrated in \cref{fig:flowchart_task2}.

\begin{enumerate}[leftmargin=2em]

\item \textbf{Experimental context and diagram enumeration.}
The agent began by dispatching, in parallel: a \textbf{\textsf{PDGSearch}} query for muon properties, a research subagent that performed 35 web searches to establish the experimental landscape (SINDRUM's measurement, Mu3e Phase~I/II sensitivity targets, HiMB projections), five \textbf{\textsf{EnumerateDiagrams}} calls for $n = 0$ through $n = 3$, and four \textbf{\textsf{EstimatePhaseSpace}} calls to compute $n$-body phase-space volumes.
The diagram counts grow super-exponentially: $1 \to 18 \to 1{,}122 \to 149{,}400$, classified into 1, 3, 5, and 7 heavy-propagator classes respectively.

\begin{figure}[t!]
\centering
\includegraphics[width=0.8\textwidth]{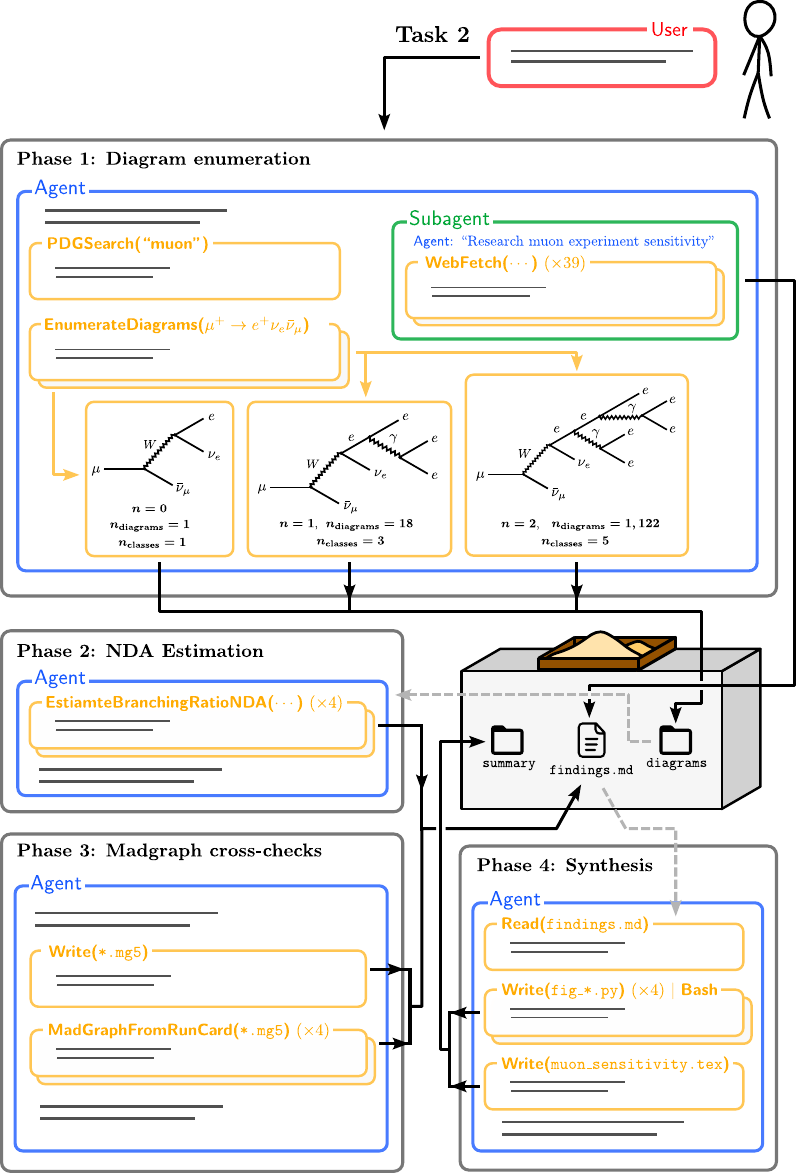}
\caption{Schematic summary flowchart of a common four-phase solution mode for the task 2 benchmark. A full transcript can be seen in appendix~\ref{app:task2_transcript}.}
\label{fig:flowchart_task2}
\end{figure}

\item \textbf{NDA estimation.}
The agent dispatched parallel \textbf{\textsf{EstimateBranchingRatioNDA}} calls for $n = 0$ through $n = 3$, passing the enumerated diagram classes by reference (directory paths to the representative diagrams).
Simultaneously, it wrote MadGraph~5 command cards with coupling-order restrictions (\texttt{QED$\leq$4} for $n=1$, \texttt{QED$\leq$6} for $n=2$) to isolate the dominant 1-$W$ topology class for exact cross-checking.

\item \textbf{MadGraph cross-checks.}
Three \textbf{\textsf{MadGraphFromRunCard}} calls validated the NDA estimates at $n = 1$ and $n = 2$.
The MadGraph results---$\text{BR}(n\!=\!1) = 3.56\times 10^{-5}$ and $\text{BR}(n\!=\!2) = 4.34\times 10^{-10}$---confirm the NDA estimates to within a factor of a few and match independent calculations and experimental measurements.

\item \textbf{Synthesis.}
The agent wrote three standalone Python plotting scripts, executed them to generate publication figures, and compiled an 8-page \LaTeX analysis document containing the full diagram classification, NDA and MadGraph comparison tables, per-pair suppression analysis, experimental frontier assessment, and a discussion of the physics driving the multiplicity scaling.

\end{enumerate}

\subsubsection{Results}
\label{sec:muon_results}

The agent's analysis document identifies the experimental frontier and provides a comprehensive breakdown of diagram counts, NDA widths, and branching ratios at each multiplicity.
\Cref{tab:scaling} summarizes the key results; the agent enumerated a total of 150{,}541 tree-level diagrams across four multiplicities.

\begin{agentresponse}
\centering
\begin{tabular}{ccccccccc}
\toprule
$n$ & Particles & \multicolumn{7}{c}{Diagrams by heavy-propagator count} \\
\cmidrule(lr){3-9}
    &           & 1$W$ & 2$W$ & 3$W$ & 4$W$ & 5$W$ & 6$W$ & 7$W$ \\
\midrule
0 & 3  & \textbf{1} & -- & -- & -- & -- & -- & -- \\
1 & 5  & \textbf{4} & 12 & 2  & -- & -- & -- & -- \\
2 & 7  & \textbf{84} & 378 & 504 & 138 & 18 & -- & -- \\
3 & 9  & \textbf{4{,}320} & 24{,}768 & 52{,}272 & 47{,}520 & 16{,}416 & 3{,}744 & 360 \\
\bottomrule
\end{tabular}
\captionof{table}{Tree-level Feynman diagram count by heavy-propagator class for $\mu^+ \to \bar\nu_\mu\nu_e\,e^+ + n(e^+e^-)$.
The dominant class (fewest heavy propagators, bold) determines the total rate at every multiplicity, with each additional heavy propagator suppressed by $(m_\mu/M_W)^4 \sim 10^{-13}$.}
\label{tab:scaling}
\end{agentresponse}

At every multiplicity, the dominant class contains exactly one heavy $W$ propagator, with all additional $e^+e^-$ pairs arising from QED bremsstrahlung.
The dominant-class fraction shrinks rapidly (100\%, 22\%, 7.5\%, 2.9\%) as subleading topologies proliferate, yet these contribute negligibly to the rate.
The MadGraph cross-checks confirm both the NDA estimates and the PDG measurement:

\begin{agentresponse}
\centering
\begin{tabular}{cccccc}
\toprule
$n$ & NDA width (GeV) & MG width (GeV) & NDA BR & MG BR & NDA/MG \\
\midrule
1 & $5.55\times10^{-24}$ & $1.066\times10^{-23}$ & $1.85\times10^{-5}$ & $3.56\times10^{-5}$ & 0.52 \\
2 & $5.65\times10^{-28}$ & $1.301\times10^{-28}$ & $1.89\times10^{-9}$ & $4.34\times10^{-10}$ & 4.3 \\
\bottomrule
\end{tabular}
\captionof{table}{Comparison of NDA estimates with exact MadGraph tree-level widths for the dominant diagram class.  The $n=1$ MadGraph result $\text{BR} = 3.56\times 10^{-5}$ agrees with the SINDRUM measurement $(3.4\pm 0.4)\times 10^{-5}$ to within one sigma.  The $n=2$ result matches the prediction of Hostert \textit{et al.}~\cite{Hostert:2023}.  NDA accuracy varies from a factor of 0.5 to 4, typical of order-of-magnitude estimation.}
\label{tab:mg_comparison}
\end{agentresponse}

\subsubsection{Multiplicity scaling and experimental frontier}
\label{sec:muon_frontier}

\begin{figure}[t]
\centering
\includegraphics[width=0.7\textwidth]{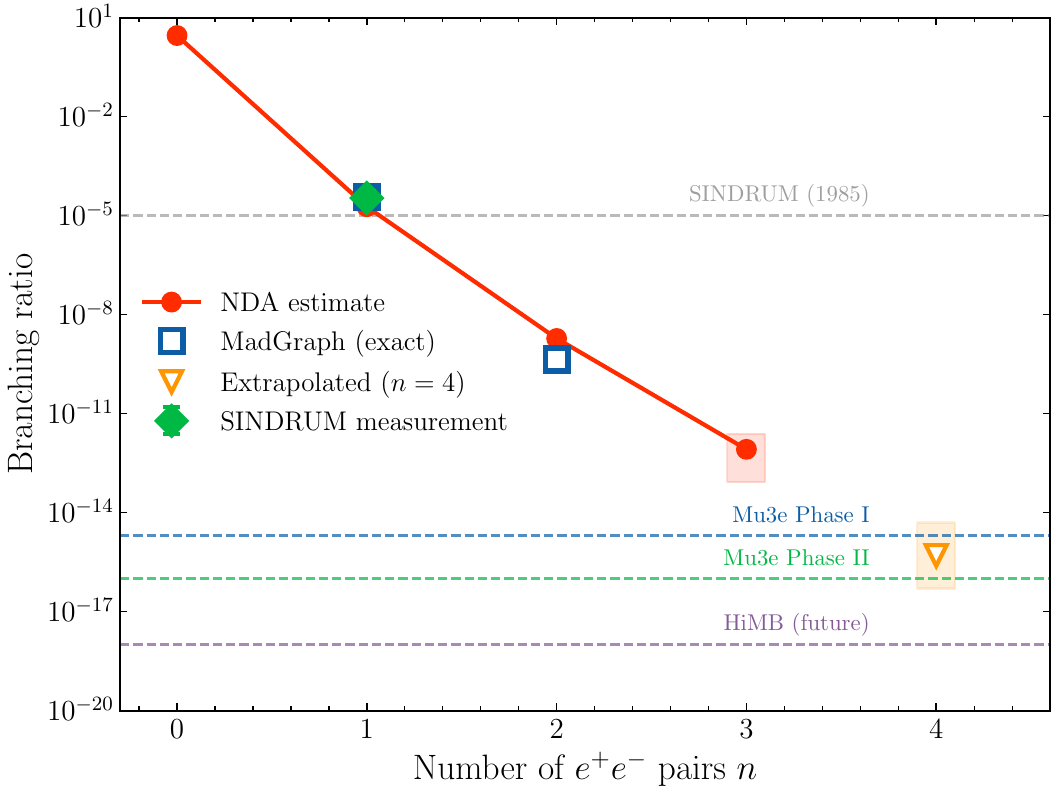}
\caption{Branching ratio of $\mu^+ \to \bar\nu_\mu \nu_e\, e^+ + n(e^+e^-)$ as a function of pair multiplicity~$n$.
NDA estimates (red circles), MadGraph exact results (blue squares), and the SINDRUM measurement (green diamond) are shown alongside experimental sensitivity thresholds.
The $n=4$ point (orange triangle) is extrapolated.  Shaded bands indicate NDA uncertainty.
\textit{Figure generated by the agent with minor cosmetic adjustments.}}
\label{fig:br_vs_n}
\end{figure}

The agent's analysis decomposes the per-pair suppression into its constituent factors:

\begin{agentresponse}
\textit{\textbf{1.\quad Per-pair suppression factor}}
\smallskip

The effective branching ratio suppression per additional $e^+e^-$ pair results from the competition of three factors:

\textbf{Coupling.} Each pair introduces two QED vertices (photon emission and $\gamma \to e^+e^-$ conversion), contributing a factor $e^4 = (4\pi\alpha)^2 \approx 8.4\times10^{-3}$ to $|\mathcal{M}|^2$.

\textbf{Phase space.} Adding two massless particles to a $k$-body final state multiplies the phase-space volume by $\Phi_{k+2}/\Phi_k$, which ranges from $6.9\times10^{-11}$ ($k = 3 \to 5$) to $2.1\times10^{-12}$ ($k = 7 \to 9$). The phase-space penalty increases with multiplicity because the available energy must be shared among more particles.

\textbf{Diagram multiplicity.} The number of dominant-class diagrams grows as $1 \to 4 \to 84 \to 4{,}320$ (ratios: $4, 21, 51$), partially compensating the coupling and phase-space suppression. Each new pair can attach its photon to any of the existing charged lines, and the identical-particle permutations multiply rapidly.

The NDA per-pair suppression weakens with each step (from $6.5\times10^{-6}$ to $4.4\times10^{-4}$) because the diagram multiplicity growth outpaces the additional phase-space penalty. The MadGraph results show a more stable suppression $\sim\!10^{-5}$ per pair, closer to the naive $(\alpha/\pi)^2 \approx 5.4\times10^{-6}$ estimate. This suggests that destructive interference among the proliferating diagrams at higher $n$ partially cancels the combinatorial enhancement that NDA assigns by summing diagram contributions incoherently.
\end{agentresponse}

\noindent The agent also identified the role of quantum interference:

\begin{figure}[t]
\centering
\includegraphics[width=0.95\textwidth]{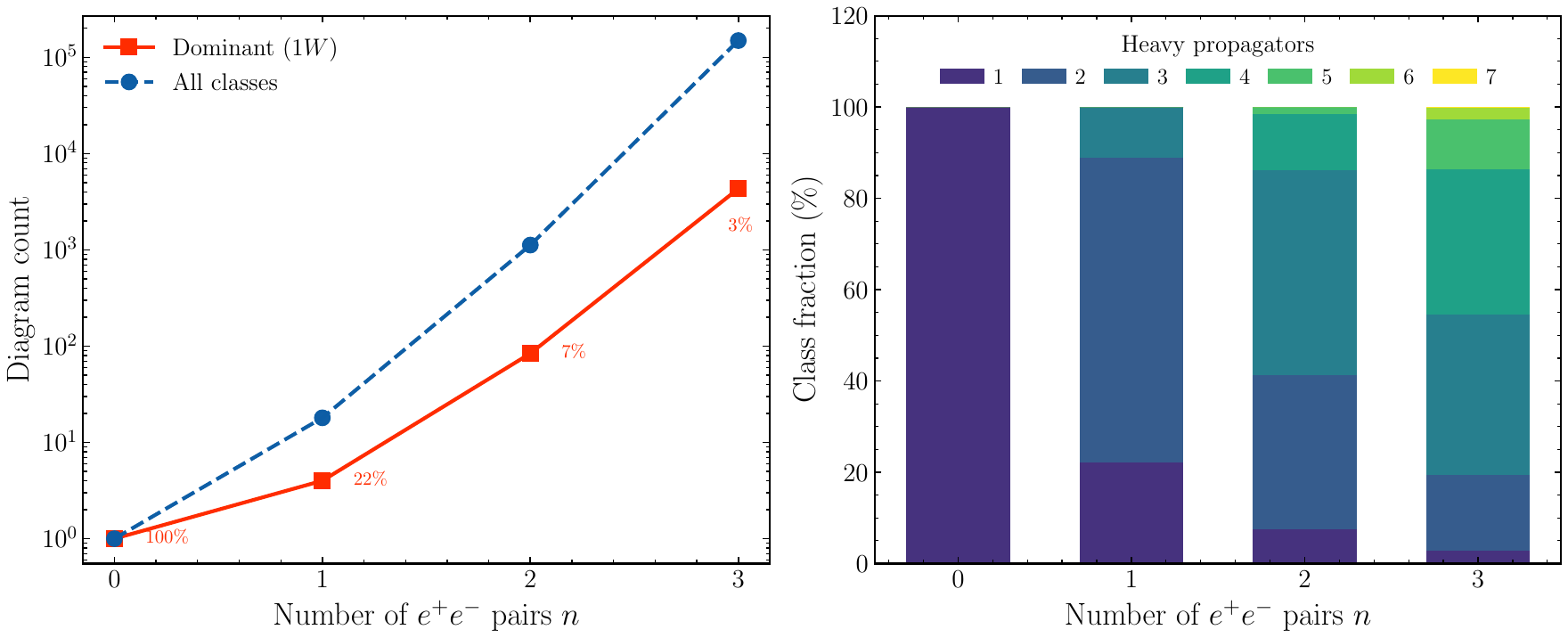}
\caption{Diagram enumeration metadata.
\textbf{Left}: dominant-class (single-$W$) count compared to total, with percentages.
\textbf{Right}: fractional composition by heavy-propagator class.
\textit{Figure generated by the agent with minor cosmetic adjustments.}}
\label{fig:diagram_metadata}
\end{figure}

\begin{agentresponse}
\textit{\textbf{2.\quad Diagram proliferation and interference}}
\smallskip

The contrast between NDA and MadGraph illuminates the role of quantum interference. NDA sums diagram contributions incoherently (adding widths), while MadGraph computes the full squared amplitude with all interference terms. At $n=1$, NDA underestimates by a factor of 2, suggesting constructive interference among the 4 dominant-class diagrams. At $n=2$, NDA overestimates by a factor of 4, indicating that destructive interference among the 84 dominant-class diagrams reduces the rate below the incoherent sum.

This transition from constructive to destructive interference as $n$ increases has a natural interpretation: at low multiplicity, the few diagrams share similar kinematic configurations and interfere constructively; at high multiplicity, the many diagrams sample a wider variety of kinematic configurations, and the random-sign phases produce partial cancellation. This effect partially counteracts the diagram multiplicity growth, stabilizing the effective per-pair suppression near $10^{-5}$.
\end{agentresponse}

\noindent The agent's assessment of the experimental frontier:

\begin{agentresponse}
\textit{\textbf{3.\quad Experimental frontier}}
\smallskip

\centering
\small
\begin{tabular}{ccccccc}
\toprule
$n$ & Final state & Total & Dominant & BR & Expected & Status \\
    &             & diags & (1$W$)  &    & events   &        \\
\midrule
0 & $e^+\nu\bar\nu$ & 1 & 1
  & $\sim\!1$
  & $2.5\times10^{15}$
  & Measured \\
1 & $e^+e^+e^-\nu\bar\nu$ & 18 & 4
  & $3.56\times10^{-5\dagger}$
  & $8.9\times10^{10}$
  & Measured \\
2 & $3e^+2e^-\nu\bar\nu$ & 1{,}122 & 84
  & $4.34\times10^{-10\dagger}$
  & $1.1\times10^{6}$
  & Mu3e Phase~I \\
3 & $4e^+3e^-\nu\bar\nu$ & 149{,}400 & 4{,}320
  & $\sim\!10^{-13}$
  & $\sim\!250$
  & Mu3e Phase~I \\
4 & $5e^+4e^-\nu\bar\nu$ & $\sim\!10^7$ & $\sim\!3\times10^5$
  & $\sim\!10^{-16}$
  & $\sim\!0.3$
  & Phase~II / HiMB \\
\bottomrule
\end{tabular}

\captionof{table}{Comprehensive summary of multi-pair muon decay.
Branching ratios marked with $\dagger$ are MadGraph exact values; others are NDA estimates.
``Expected events'' assumes the full Mu3e Phase~I dataset ($2.5\times10^{15}$ muons).}
\label{tab:frontier}

\smallskip
\raggedright

\textbf{$n = 2$ (first observation).} With $\text{BR} = 4.34\times10^{-10}$ and $2.5\times10^{15}$ muon stops at Mu3e Phase~I, approximately $10^6$ signal events are expected --- a precision measurement, not just a discovery. The five charged tracks in the final state provide a distinctive experimental signature.
\vspace{0.1in}

\textbf{$n = 3$ (frontier).} The NDA estimate gives $\text{BR} \sim 8\times10^{-13}$, which based on the NDA/MG calibration at lower multiplicities likely corresponds to a true branching ratio in the range $10^{-14}$--$10^{-12}$. Even at the conservative end, this yields $\sim\!25$--$2{,}500$ events at Mu3e Phase~I, making $n=3$ observable with seven charged tracks. The event yield grows by a factor of $\sim\!20$ at Phase~II.
\vspace{0.1in}

\textbf{$n = 4$ (beyond the frontier).} Extrapolating the per-pair suppression, we estimate $\text{BR}(n\!=\!4) \sim 10^{-17}$--$10^{-16}$. The lower end lies below Mu3e Phase~II sensitivity ($\text{SES} \sim 10^{-16}$) but could potentially be accessible with the full HiMB upgrade ($\text{SES} \sim 10^{-18}$). The nine charged tracks would provide an unmistakable signature.
\vspace{0.1in}

\textbf{$n \geq 5$.} At $\text{BR} \lesssim 10^{-20}$, the $n = 5$ channel is far below any foreseeable muon experiment. We identify $n = 4$ as the absolute upper limit of observability.
\end{agentresponse}

\noindent These results demonstrate the complementary use of NDA for rapid triage and MadGraph for precision validation within a single agentic session.
The agent autonomously identified the hierarchy of diagram classes, recognized that subleading topologies are suppressed by 13 orders of magnitude per additional heavy propagator, and mapped the experimental frontier from the interplay of coupling suppression, phase-space compression, and combinatorial growth.

\section{Conclusions}
\label{sec:conclusions}

In this work, we introduced \textit{\adname}, a symbolic computation extension to the \pkgname agentic toolkit that enables LLM agents to plan and execute multi-step theoretical calculations in high-energy physics.
The conceptual backbone of the architecture is an entropy decomposition of the agent's action distribution into task, context, and execution components ($\Delta_T$, $\Delta_C$, $\Delta_E$), which identifies execution uncertainty as the primary obstacle to reliable agentic symbolic computation.
Unlike traditional programming, where compilers and runtime checks can at least enforce syntactic and type correctness, symbolic computation is governed by implicit mathematical conventions that no computational backend can efficiently verify independently.
We argued that this challenge admits two complementary remedies, tool-constrained computation that drives $\Delta_E$ toward zero structurally and targeted knowledge grounding that suppresses it probabilistically, and pursued the first as the primary architecture.

The toolkit is organized around a shared LLM-friendly diagram specification that serves as a single source of truth across two complementary calculation paths.
Naive Dimensional Analysis provides order-of-magnitude rate estimates applicable to arbitrary final-state multiplicities, while Exact Diagrammatic Analysis produces complete tree-level symbolic results through automatic \fc code generation.
Both paths consume the same structured JSON specification and are fed by \fg for automatic Feynman diagram enumeration, with a navigable theory knowledge base exposed through the same tool interface.
By treating diagram specifications, symbolic results, and generated scripts as persistent, referenceable artifacts within the agent's orchestration loop, the architecture preserves transparency, provenance, and human oversight throughout multi-step workflows.

Two benchmarks, each completed autonomously from a single user prompt, validated the architecture.
Task~1 exercised the \EDA path on an exhaustive catalog of all tree-level, single-vertex $1\to 2$ decay widths, producing 19 independent formulas across 6 vertex families with complete massless and threshold limits, chiral cross-checks, and Standard Model validation at the few-percent level.
Task~2 exercised the NDA path and \fg on the muon decay $\mu \to \nu_\mu\bar{\nu}_e + n(e^+e^-) + e^-$, enumerating over 150{,}000 tree-level diagrams across four multiplicities to map the experimental frontier from established measurements through the reach of Mu3e and HiMB, with MadGraph cross-checks confirming both the NDA scaling and published data. 
Both tasks are specified with sufficient detail to serve as reproducible test cases for other agentic platforms, and we encourage other groups to attempt them and report their results.

Several design principles emerged from this work:
\begin{enumerate}
    \item The reliability of agentic symbolic computation is determined less by the model's familiarity with domain conventions than by whether the tool interface encodes them. Concentrating convention-dependent choices into schema-validated fields eliminates entire classes of silent errors by construction, independently of model capability.
    \item A multi-fidelity organization, in which approximate and exact calculations consume the same structured specification, lets the agent match computational cost to the question being asked while providing internal consistency checks.
    \item Treating generated scripts, symbolic results, and intermediate outputs as persistent, referenceable artifacts rather than transient context allows downstream tools to consume results by reference, avoiding the context-window inflation that would otherwise degrade attention and retrieval fidelity over multi-step pipelines.
    \item Co-locating procedural tools and theoretical reference material in the same tool interface allows the agent to look up conventions at the moment of each critical decision, suppressing execution uncertainty through targeted retrieval rather than bulk documentation.

\end{enumerate}

These principles point to a broader observation about the role of tools in agentic scientific computation.
The growing literature on LLM agents in HEP~\cite{LLMAgentSurvey,AIScience,MadAgents,CoLLM,ColliderAgent,HEPDataAnalysisAgents,GRACE,Agrawal:2026fermiacc,Tan:2026physmaster,ArgoLOOM,Moreno:2026mqk} has predominantly adopted a \emph{code generation} paradigm: the agent writes free-form scripts or shell commands that are executed against existing software, with the reliability burden falling entirely on the model's ability to produce correct code from context.
This approach inherits the full execution uncertainty $\Delta_E$ analyzed in \cref{sec:reliability}, since the agent must independently resolve every domain-dependent choice at the token level.
The \pkgname framework~\cite{Menzo:2025cim} was designed from the outset around schema-validated tool invocation rather than code generation, and \adname continues this approach into the domain of symbolic computation, where the stakes are particularly high because correctness is governed by implicit mathematical conventions rather than compiler-enforceable rules.
The key is to design tool schemas whose valid completions fix not only the format but the \emph{domain conventions} of the calculation, so that every schema-compliant invocation maps to a correct computation by construction.
This yields two distinct advantages.
The first is \emph{reliability}: correctness of the mechanical steps becomes a property of the interface rather than of the model, eliminating entire classes of silent errors independently of model capability.
The second is \emph{interpretability}: because the agent's decisions are compressed into a small number of human-readable fields, a domain expert can audit the full specification at a glance, in sharp contrast to reviewing pages of generated code where errors may be invisible without re-deriving the calculation from scratch.
We believe convention-fixing tool design, in which the structure of the action space encodes as much domain knowledge as the model that navigates it, deserves wider adoption in agentic HEP workflows.

The present architecture suggests several natural directions for future development.
Perhaps the most natural next step is to seamlessly integrate \adname's symbolic outputs to \pkgname's existing Monte Carlo tools (MadGraph~\cite{MadGraph}, Pythia~\cite{Pythia}, Sherpa~\cite{Sherpa}), which would close the loop between symbolic theory and simulated events within a single agentic session and realize the full potential of the \pkgname framework as an end-to-end research tool.
More specifically, the NDA path can be extended to more refined power-counting prescriptions, such as generalized NDA rules~\cite{Gavela:2016bzc} for arbitrary EFTs or warped/composite NDA for strongly coupled scenarios~\cite{Panico:2015jxa}, with the tool interface designed to accommodate alternative scaling prescriptions without changing the diagram specification.
For the \EDA path, three extensions would significantly broaden the toolkit's scope: one-loop calculations via Package-X~\cite{PackageX,PackageX2} for scalar integral evaluation, flavor-indexed coupling matrices enabling decay widths with full interference structure for non-diagonal flavor couplings, and automatic parsing of UFO model files from FeynRules~\cite{FeynRules} into \fg vertex rules for BSM diagram enumeration without manual model specification.
More broadly, the tool-constrained and knowledge-grounded approaches discussed in \cref{sec:tools_vs_knowledge} suggest a trajectory for increasingly autonomous systems that begins with the tool-constrained architecture presented here and progressively expands the boundary of free-form reasoning as the knowledge base deepens and verification tools mature.
Multi-agent configurations, where specialized agents for symbolic calculation, Monte Carlo validation, and data comparison are coordinated by a planning agent, represent a natural evolution toward this goal.

\adname provides the symbolic computation layer required for increasingly capable AI-assisted research in high-energy physics.
The design principles identified here, including convention-fixing semantics, multi-fidelity tool hierarchies, shared structured specifications, and theory-grounded interfaces, offer a blueprint for reliable agentic computation that can extend a researcher's effective capabilities beyond what either the researcher or the agent could achieve working independently.

\section{Public code and data availability}
\label{sec:data_availability}

The open source code for this project, including all \adname tools, the theory knowledge base, system prompts, and benchmark session logs can be found in the HEPTAPOD repository:
\begin{center}
\url{https://github.com/tonymenzo/heptapod}
\end{center}
The repository includes the unedited Claude Code conversation logs in JSONL format, agent-generated deliverables (compiled reference documents, figures, and analysis scripts), as well as a conversation analysis utility used to produce the curated transcript excerpts in appendices~\ref{app:task1_transcript}, \ref{app:task2_transcript} and the session statistics reported in \cref{sec:task1_workflow,sec:task2_workflow}.

\section*{Acknowledgments}

TM is supported in part by the Shelby Endowment for Distinguished Faculty at the University of Alabama and by Fermilab via Subcontract 725339. 
The work of AR and KM is supported in part by the Shelby Endowment for Distinguished Faculty at the University of Alabama and by Fermilab via Subcontract 731293, in support of DOE Award No.\ DE-SCL0000090 ``HEP AmSC IDA Pilot: Knowledge Extraction'' and DOE Award No.\ DE-SCL0000152 ``USQCD AmSC Infrastructure Provision''. The work of SG and KM is supported in part by the U.S. Department of Energy (DOE) under Award No. DE-SC0026347. The work of SG is supported in part by DOE under Award No. DE-SC0012447. 
This manuscript has been authored by Fermi Forward Discovery Group, LLC under Contract No.~89243024CSC000002 with the U.S. Department of Energy, Office of Science, Office of High Energy Physics.

\bibliographystyle{JHEP}
\bibliography{ref}

\newpage

\appendix
\section{NDA prescription for decay rates and cross sections}
\label{app:nda_prescription}

This appendix summarizes the explicit NDA prescription implemented by the toolkit. See also Ref.~\cite{Alwall:2014bza}.

\subsection{Exact starting point}

For a $1\to n$ decay of a mother particle $A$ with mass $M$,
\begin{equation}
    \Gamma(A \to a_1 a_2 \cdots a_n)
    =
    \frac{1}{2M}
    \int d\Phi_n \,
    \overline{|\mathcal{M}|^2}.
    \label{eq:app_decay_def}
\end{equation}
For a $2\to n$ scattering process in the center-of-mass frame,
\begin{equation}
    \sigma(AB \to a_1 a_2 \cdots a_n)
    =
    \frac{1}{F}
    \int d\Phi_n \,
    \overline{|\mathcal{M}|^2},
    \label{eq:app_xs_def}
\end{equation}
with flux factor
\begin{equation}
    F = 4E_A E_B |\vec v_A - \vec v_B|.
\end{equation}
In the center-of-mass frame with relativistic incoming particles, $E_A = E_B = \sqrt{s}/2$ and $F \to 2s$.

\subsection{Phase-space estimates}

For two-body decays, the toolkit uses the exact phase-space factor
\begin{equation}
    \Phi_2 = \frac{|\vec p\,|}{8\pi M},
\end{equation}
where $|\vec p\,|$ is the daughter three-momentum in the parent rest frame.
For $n$-body final states with approximately massless daughters, the phase-space volume is estimated by
\begin{equation}
    \Phi_n(M)
    =
    \frac{1}{2(4\pi)^{2n-3}}
    \frac{M^{2n-4}}{\Gamma(n)\Gamma(n-1)}.
    \label{eq:app_phasespace_decay}
\end{equation}
For $2\to n$ scattering in the center-of-mass frame, the same massless phase-space scaling gives
\begin{equation}
    \Phi_n(\sqrt{s})
    \propto s^{\,n-2}.
    \label{eq:app_phasespace_scatt}
\end{equation}
These forms are adequate for NDA purposes, but should not be interpreted as precision approximations near threshold or when several final-state masses are comparable to the available energy.

\subsection{Matrix element factorization}

The squared matrix element is estimated as a modular product of diagram components,
\begin{equation}
    \overline{|\mathcal{M}|^2}_{\text{NDA}}
    \sim
    \frac{N_{\text{color}}}{N_{\text{spin}}}
    \left(
        \prod_{\text{vertices}} |C_i|^2
    \right)
    \left(
        \prod_{\text{internal}} |P_i(E)|^2
    \right)
    \left(
        \prod_{\text{external}} \mathcal{P}_i(E)
    \right)
    \left(
        \frac{1}{16\pi^2}
    \right)^L,
    \label{eq:app_matrix_factorization}
\end{equation}
where:
\begin{itemize}[leftmargin=2em]
    \item $N_{\text{spin}}$ denotes the average over initial-state spin multiplicities;
    \item $N_{\text{color}}$ denotes the color multiplicity of the diagram;
    \item $C_i$ are the vertex couplings or operator coefficients;
    \item $P_i(E)$ are propagator suppression factors evaluated at a characteristic scale $E$;
    \item $\mathcal{P}_i(E)$ are rough external-state polarization or spinor factors;
    \item $L$ is the loop order.
\end{itemize}
The characteristic scale is taken to be $E \sim M$ for decays and $E \sim \sqrt{s}$ for scattering in the center-of-mass frame.

\subsection{External-state factors}

The rough external-state factors used by the toolkit are chosen to reproduce the energy scaling of the standard spin sums and polarization sums:
\begin{equation}
    \mathcal{P}_{s=0}(E) \sim 1,
    \qquad
    \mathcal{P}_{s=1/2}(E) \sim 2E,
    \qquad
    \mathcal{P}_{s=1}(E) \sim
    \begin{cases}
        1 + E^2/m^2, & m \neq 0, \\
        1, & m = 0.
    \end{cases}
    \label{eq:app_external_factors}
\end{equation}
These factors should be interpreted as scaling surrogates for the exact tensor and spinor structures.
They are not intended to reproduce chiral suppressions, gauge cancellations, or interference effects.

\subsection{Propagator regimes}

For an internal propagator with mass $m_{\text{prop}}$, the toolkit distinguishes three regimes:
\begin{itemize}[leftmargin=2em]
    \item \textbf{heavy}: $m_{\text{prop}} \gg E$, giving 
    $|P_i|^2 \sim 1/m_{\text{prop}}^4$;
    \item \textbf{light}: $m_{\text{prop}} \ll E$, giving $|P_i|^2 \sim 1/E^4$;
    \item \textbf{intermediate}: use the full scaling 
    $|P_i|^2 \sim 1/|q^2-m_{\text{prop}}^2|^2$ at the characteristic kinematic scale.
\end{itemize}
These suppressions are applied diagram by diagram.
They should not be used when the resolved propagator is better described by an effective local operator; in that case the operator coefficient already encodes the heavy-scale suppression.

\subsection{Higher-dimensional operators}

For an operator of mass dimension $d>4$ with coefficient $C^{(d)}/\Lambda^{d-4}$, the amplitude scales as
\begin{equation}
    \mathcal{M}^{(d)}
    \sim
    C^{(d)} E^{\,d-N}\Lambda^{\,4-d},
    \label{eq:app_hdo_amp}
\end{equation}
where $N$ is the total number of external particles in the process.
This reproduces the correct mass dimension $4-N$ for the amplitude.
For local contact interactions of this type, it is often more transparent to estimate the energy scaling directly from \cref{eq:app_hdo_amp} rather than separately keeping resolved propagator denominators.

\subsection{Final NDA forms}

With these ingredients, the decay width estimate becomes
\begin{equation}
    \Gamma_{\text{NDA}}
    \sim
    \frac{1}{2M}
    \Phi_n(M)
    \overline{|\mathcal{M}|^2}_{\text{NDA}},
    \label{eq:app_final_nda_decay}
\end{equation}
and for scattering in the center-of-mass frame,
\begin{equation}
    \sigma_{\text{NDA}}
    \sim
    \frac{1}{2s}
    \Phi_n(\sqrt{s})
    \overline{|\mathcal{M}|^2}_{\text{NDA}}.
    \label{eq:app_final_nda_scatt}
\end{equation}
The two-body forms used in the main text are recovered as special cases of these expressions.

\section{Complete \texorpdfstring{$1\to 2$}{1 -> 2} decay width formulas}
\label{app:catalog}

This appendix reproduces the decay width formulas from the agent's reference document (Task~1), organized by parent spin.
All widths are spin-averaged over the initial state,
\begin{equation}
    \Gamma = \frac{1}{2s_P+1}\,\frac{|\mathbf{p}|}{8\pi M^2}\,\overline{|\mathcal{M}|^2}
\end{equation}
where $|\mathbf{p}| = \sqrt{\lambda(M^2,m_1^2,m_2^2)}/(2M)$ is the center-of-mass momentum and
\begin{equation}
\lambda(a,b,c) = a^2+b^2+c^2-2ab-2bc-2ca
\end{equation}
is the K\"all\'en function.
Couplings are complex throughout; $|g|^2 \equiv g\,g^*$.
We abbreviate $\lambda_{12} \equiv \lambda(M^2,m_1^2,m_2^2)$.
The basis changes connecting dual parameterizations are:
\begin{equation}
\text{SFF:}\quad g_S = \frac{g_L+g_R}{2}\,,\quad g_P = \frac{g_R-g_L}{2}\,;\qquad
\text{VFF:}\quad g_V = \frac{g_L+g_R}{2}\,,\quad g_A = \frac{g_L-g_R}{2}\,.
\end{equation}


\begin{agentresponse}
{\large\textit{\textbf{1.\quad Spin-0 parent}}}
\medskip

\paragraph{$S \to f\bar{f}$, scalar--pseudoscalar basis.}
\begin{equation}
\boxed{
\Gamma(S\to f\bar{f}) = \frac{\sqrt{\lambda(m_S^2, m_f^2, m_{\bar f}^2)}}{8\pi\, m_S^3}
\left[|g_S|^2\bigl(m_S^2 - (m_f+m_{\bar f})^2\bigr)
+ |g_P|^2\bigl(m_S^2 - (m_f-m_{\bar f})^2\bigr)\right]
}
\label{eq:S-ff-va}
\end{equation}

The scalar coupling $g_S$ produces a P-wave ($\beta^3$) threshold for equal-mass fermions,
while the pseudoscalar coupling $g_P$ produces an S-wave ($\beta^1$) threshold.
The interference between $g_S$ and $g_P$ vanishes identically in the spin-summed amplitude.

\paragraph{Chiral basis.}
\begin{equation}
\Gamma(S\to f\bar{f}) = \frac{\sqrt{\lambda(m_S^2, m_f^2, m_{\bar f}^2)}}{16\pi\, m_S^3}
\left[(|g_L|^2+|g_R|^2)(m_S^2-m_f^2-m_{\bar f}^2)
- 2\,\mathrm{Re}(g_L g_R^*)\,2m_f m_{\bar f}\right]
\label{eq:S-ff-ch}
\end{equation}

\paragraph{Special limits.}

\noindent\textit{Equal mass} ($m_{\bar f} = m_f$):
\begin{equation}
\Gamma = \frac{\sqrt{m_S^2-4m_f^2}}{8\pi\, m_S^2}
\left[|g_P|^2\, m_S^2 + |g_S|^2\,(m_S^2-4m_f^2)\right]
\label{eq:S-ff-va-equal}
\end{equation}

\noindent\textit{Massless fermions} ($m_f = m_{\bar f} = 0$):
\begin{equation}
\Gamma = \frac{m_S}{8\pi}\,(|g_S|^2+|g_P|^2)
= \frac{m_S}{16\pi}\,(|g_L|^2+|g_R|^2)
\label{eq:S-ff-massless}
\end{equation}

\paragraph{$S \to S_1 S_2$ (SSS).}
\begin{equation}
\boxed{
\Gamma(S\to S_1 S_2) = \frac{|g|^2}{16\pi\, m_S^3}\,
\sqrt{\lambda(m_S^2,\, m_{S_1}^2,\, m_{S_2}^2)}
}
\label{eq:SSS}
\end{equation}

This is the simplest decay width formula: pure phase space times a coupling-squared.
The threshold behavior is $\Gamma \propto \beta$ (S-wave) for equal-mass daughters.

\paragraph{Equal mass} ($m_{S_2} = m_{S_1}$):
$\Gamma = |g|^2 \sqrt{m_S^2 - 4m_{S_1}^2}\,/\,(16\pi\, m_S^2)$.

\paragraph{$S \to V_1 V_2$ --- SVV vertices.}

Three Lorentz structures are available for scalar $\to$ vector pair:

\subparagraph{Renormalizable SVV ($g\,\varepsilon_1\cdot\varepsilon_2$).}
\begin{equation}
\boxed{
\Gamma = \frac{|g|^2\,\sqrt{\lambda_S}}{16\pi\, m_S^3}
\left[2 + \frac{(m_{V_1}^2+m_{V_2}^2-m_S^2)^2}{4m_{V_1}^2 m_{V_2}^2}\right]
}
\label{eq:SVV}
\end{equation}
where $\lambda_S \equiv \lambda(m_S^2, m_{V_1}^2, m_{V_2}^2)$.
The term $2$ counts the two transverse polarization pairs; the second term arises from
longitudinal modes and grows as $m_S^4/(4m_{V_1}^2 m_{V_2}^2)$ for $m_S \gg m_V$,
reflecting the Goldstone boson equivalence theorem.

\subparagraph{Dim-5 field strength ($\phi FF$).}
\begin{equation}
\boxed{
\Gamma = \frac{|g|^2\,\sqrt{\lambda_S}}{8\pi\, m_S^3}
\bigl(m_S^4 - 2m_S^2(m_{V_1}^2+m_{V_2}^2) + m_{V_1}^4+4m_{V_1}^2 m_{V_2}^2+m_{V_2}^4\bigr)
}
\label{eq:dim5FF}
\end{equation}

\subparagraph{Dim-5 dual field strength ($\phi F\tilde{F}$).}
\begin{equation}
\boxed{
\Gamma = \frac{|g|^2}{8\pi\, m_S^3}\,\lambda_S^{3/2}
}
\label{eq:dim5FFdual}
\end{equation}

The $\phi F\tilde{F}$ operator produces a $\beta^3$ (P-wave) threshold, in contrast to the
$\beta^1$ threshold of both the renormalizable SVV and $\phi FF$ operators.
This is because the Levi-Civita tensor contracts two momenta, introducing an extra factor
of $|\mathbf{p}|^2 \propto \beta^2$ near threshold.

\paragraph{Equal-mass limits} ($m_{V_2}=m_{V_1}\equiv m_V$):
\begin{align}
\Gamma_{\rm SVV} &= \frac{|g|^2\sqrt{m_S^2-4m_V^2}}{64\pi\, m_S^2 m_V^4}
\bigl(m_S^4-4m_S^2 m_V^2+12m_V^4\bigr) \label{eq:SVV-equal}\\[4pt]
\Gamma_{\phi FF} &= \frac{|g|^2\sqrt{m_S^2-4m_V^2}}{8\pi\, m_S^2}
\bigl(m_S^4-4m_S^2 m_V^2+6m_V^4\bigr) \label{eq:FF-equal}\\[4pt]
\Gamma_{\phi F\tilde{F}} &= \frac{|g|^2}{8\pi}\,(m_S^2-4m_V^2)^{3/2}
\label{eq:FFd-equal}
\end{align}
\end{agentresponse}

\begin{agentresponse}
{\large\textit{\textbf{2.\quad Spin-$\frac{1}{2}$ parent}}}
\medskip

\paragraph{$f_1 \to S\,\bar{f}_2$, scalar--pseudoscalar basis.}
\begin{equation}
\boxed{
\Gamma(f_1\to S\,\bar{f}_2) = \frac{\sqrt{\lambda(m_{f_1}^2, m_S^2, m_{f_2}^2)}}{16\pi\, m_{f_1}^3}
\left[|g_S|^2\bigl((m_{f_1}-m_{f_2})^2-m_S^2\bigr)
+ |g_P|^2\bigl((m_{f_1}+m_{f_2})^2-m_S^2\bigr)\right]
}
\label{eq:f-Sf-va}
\end{equation}

Note the swapped roles of $g_S$ and $g_P$ relative to \eqref{eq:S-ff-va}: the scalar coupling
produces S-wave threshold here (factor $(m_{f_1}-m_{f_2})^2 - m_S^2 \to m_{f_1}^2$ for massless daughters), while
the pseudoscalar coupling gets an enhancement from the $(m_{f_1}+m_{f_2})^2$ numerator.

\paragraph{Chiral basis.}
\begin{equation}
\Gamma = \frac{\sqrt{\lambda(m_{f_1}^2, m_S^2, m_{f_2}^2)}}{32\pi\, m_{f_1}^3}
\left[(|g_L|^2+|g_R|^2)(m_{f_1}^2+m_{f_2}^2-m_S^2)
- 2\,\mathrm{Re}(g_L g_R^*)\cdot 2m_{f_1}m_{f_2}\right]
\label{eq:f-Sf-ch}
\end{equation}

\paragraph{$f_1 \to V\,\bar{f}_2$, vector--axial basis.}
\begin{equation}
\boxed{
\Gamma = \frac{\sqrt{\lambda_f}}{16\pi\, m_{f_1}^3 m_V^2}
\left[|g_V|^2\,\mathcal{P}_+(m_{f_1},m_{f_2},m_V)
+ |g_A|^2\,\mathcal{P}_-(m_{f_1},m_{f_2},m_V)\right]
}
\label{eq:f-Vf-va}
\end{equation}
where $\lambda_f \equiv \lambda(m_{f_1}^2, m_V^2, m_{f_2}^2)$ and
\begin{equation}
\mathcal{P}_\pm(M,m,m_V) = M^4+m^4+m^2 m_V^2-2m_V^4+M^2(m_V^2-2m^2) \pm 6Mm\,m_V^2\,.
\end{equation}

The $\pm 6Mm\,m_V^2$ term arises from the interference between transverse and longitudinal polarizations
with the fermion mass insertion. It distinguishes vector from axial couplings and vanishes for
massless daughter fermions.

\paragraph{Chiral basis.}
\begin{equation}
\Gamma = \frac{\sqrt{\lambda_f}}{32\pi\, m_{f_1}^3 m_V^2}
\left[(|g_L|^2+|g_R|^2)\,\mathcal{P}_0 + 6\,\mathrm{Re}(g_L g_R^*)\,m_{f_1} m_{f_2} m_V^2\right]
\label{eq:f-Vf-ch}
\end{equation}
where $\mathcal{P}_0 = \mathcal{P}_+|_{6Mm\to 0} = M^4+m^4+m^2 m_V^2-2m_V^4+M^2(m_V^2-2m^2)$.

\paragraph{Massless daughter} ($m_{f_2}=0$):
\begin{equation}
\Gamma = \frac{(m_{f_1}^2-m_V^2)^2(m_{f_1}^2+2m_V^2)}{16\pi\, m_{f_1}^3 m_V^2}\,(|g_V|^2+|g_A|^2)
\label{eq:f-Vf-massless}
\end{equation}

\paragraph{Tensor (dipole) vertex.}
The tensor coupling $ig\,\sigma^{\mu\nu}k_\nu$ gives:
\begin{equation}
\boxed{
\Gamma = \frac{|g|^2\,\sqrt{\lambda_f}}{16\pi\, m_{f_1}^3}
\left(2m_{f_1}^4+2m_{f_2}^4+6m_{f_1} m_{f_2} m_V^2-m_{f_2}^2 m_V^2-m_V^4-m_{f_1}^2(4m_{f_2}^2+m_V^2)\right)
}
\label{eq:f-Vf-tensor}
\end{equation}

\paragraph{Tensor-chiral vertex.}
$i(g_L P_L + g_R P_R)\sigma^{\mu\nu}k_\nu$:
\begin{equation}
\Gamma = \frac{\sqrt{\lambda_f}}{32\pi\, m_{f_1}^3}
\left[(|g_L|^2+|g_R|^2)\,\mathcal{T}_0 + 6\,\mathrm{Re}(g_L g_R^*)\,m_{f_1} m_{f_2} m_V^2\right]
\label{eq:f-Vf-tch}
\end{equation}
where $\mathcal{T}_0 = 2m_{f_1}^4+2m_{f_2}^4-m_{f_2}^2 m_V^2-m_V^4-m_{f_1}^2(4m_{f_2}^2+m_V^2)$.

Setting $g_L=g_R=g$ in \eqref{eq:f-Vf-tch} reproduces \eqref{eq:f-Vf-tensor}, confirming consistency.
\end{agentresponse}

\begin{agentresponse}
{\large\textit{\textbf{3.\quad Spin-1 parent}}}
\medskip

The spin-averaging factor $1/(2s_P+1) = 1/3$ for vector parents introduces an overall factor of
$1/3$ relative to the scalar-parent formulas.

\paragraph{$V \to f\bar{f}$, vector--axial basis.}
Define $Q_V \equiv 2m_V^4-m_V^2(m_f^2+m_{\bar f}^2)-(m_f^2-m_{\bar f}^2)^2$.
\begin{equation}
\boxed{
\Gamma(V\to f\bar{f}) = \frac{\sqrt{\lambda(m_V^2,m_f^2,m_{\bar{f}}^2)}}{24\pi\, m_V^5}
\left[|g_V|^2\bigl(Q_V+6m_f m_{\bar f}m_V^2\bigr)
+|g_A|^2\bigl(Q_V-6m_f m_{\bar f}m_V^2\bigr)\right]
}
\label{eq:V-ff-va}
\end{equation}

\paragraph{Chiral basis.}
\begin{equation}
\Gamma = \frac{\sqrt{\lambda_V}}{48\pi\, m_V^5}
\left[(|g_L|^2+|g_R|^2)\,Q_V
+6\,\mathrm{Re}(g_L g_R^*)\,m_f m_{\bar f} m_V^2\right]
\label{eq:V-ff-ch}
\end{equation}

\paragraph{Equal mass} ($m_{\bar f}=m_f$):
\begin{equation}
\Gamma = \frac{\sqrt{m_V^2-4m_f^2}}{12\pi\, m_V^2}
\left[|g_V|^2(2m_f^2+m_V^2) + |g_A|^2(m_V^2-4m_f^2)\right]
\label{eq:V-ff-va-equal}
\end{equation}

\paragraph{Massless fermions} ($m_f=m_{\bar f}=0$):
\begin{equation}
\Gamma = \frac{m_V}{12\pi}\,(|g_V|^2+|g_A|^2) = \frac{m_V}{24\pi}\,(|g_L|^2+|g_R|^2)
\label{eq:V-ff-massless}
\end{equation}

\paragraph{Tensor (dipole) vertex.}
\begin{equation}
\boxed{
\Gamma = \frac{|g|^2\,\sqrt{\lambda_V}}{24\pi\, m_V^3}
\bigl(-2m_f^4+m_f^2(4m_{\bar f}^2+m_V^2)+6m_f m_{\bar f} m_V^2-2m_{\bar f}^4+m_{\bar f}^2 m_V^2+m_V^4\bigr)
}
\label{eq:V-ff-tensor}
\end{equation}

Equal mass: $\Gamma = |g|^2\sqrt{m_V^2-4m_f^2}\,(8m_f^2+m_V^2)/(24\pi)$.
Massless: $\Gamma = |g|^2 m_V^3/(24\pi)$ --- note the $m_V^3$ scaling (vs.\ $m_V$ for the gauge vertex),
a hallmark of the dimension-5 operator.

\paragraph{Tensor-chiral vertex.}
\begin{equation}
\Gamma = \frac{\sqrt{\lambda_V}}{48\pi\, m_V^3}
\left[(|g_L|^2+|g_R|^2)\mathcal{T}_0^V
+ 6\,\mathrm{Re}(g_L g_R^*)\,m_f m_{\bar f} m_V^2\right]
\label{eq:V-ff-tch}
\end{equation}
where $\mathcal{T}_0^V = -2m_f^4+m_f^2(4m_{\bar f}^2+m_V^2)-2m_{\bar f}^4+m_{\bar f}^2 m_V^2+m_V^4$.

\paragraph{$V \to S_1 S_2$ (SSV).}
\begin{equation}
\boxed{
\Gamma(V\to S_1 S_2) = \frac{|g|^2}{48\pi\, m_V^5}\,\lambda(m_V^2, m_{S_1}^2, m_{S_2}^2)^{3/2}
}
\label{eq:SSV}
\end{equation}

The $\lambda^{3/2}$ factor produces a $\beta^3$ (P-wave) threshold, because the SSV vertex couples
through the momentum difference $(p_1-p_2)^\mu$, contributing an extra power of $|\mathbf{p}|$.
Equal mass: $\Gamma = |g|^2(m_V^2-4m_{S}^2)^{3/2}/(48\pi\, m_V^2)$.

\paragraph{$V \to S\,V'$ (SVV, vector parent).}
\begin{equation}
\boxed{
\Gamma(V\to S\,V') = \frac{|g|^2\,\sqrt{\lambda(m_V^2,m_S^2,m_{V'}^2)}}{48\pi\, m_V^3}
\left[2 + \frac{(m_V^2+m_{V'}^2-m_S^2)^2}{4m_V^2 m_{V'}^2}\right]
}
\label{eq:V-SV}
\end{equation}

This has the same polarization structure as the scalar-parent SVV decay \eqref{eq:SVV},
with an additional $1/3$ from spin averaging.

\paragraph{$V \to V_1 V_2$ (VVV).}
\begin{equation}
\boxed{
\Gamma(V\to V_1 V_2) = \frac{|g|^2\,\sqrt{\lambda_V}}{192\pi\, m_V^5\, m_{V_1}^2 m_{V_2}^2}\,\mathcal{Q}(m_V, m_{V_1}, m_{V_2})
}
\label{eq:VVV}
\end{equation}
where
\begin{align}
\mathcal{Q} &= m_V^8 - 4m_V^6(m_{V_1}^2+m_{V_2}^2) + 2m_V^4(3m_{V_1}^4+8m_{V_1}^2 m_{V_2}^2+3m_{V_2}^4) \nonumber\\
&\quad - 4m_V^2(m_{V_1}^6+5m_{V_1}^4 m_{V_2}^2+5m_{V_1}^2 m_{V_2}^4+m_{V_2}^6) \nonumber\\
&\quad + (m_{V_1}^2-m_{V_2}^2)^2(m_{V_1}^4+10m_{V_1}^2 m_{V_2}^2+m_{V_2}^4)\,.
\end{align}

Equal mass ($m_{V_2}=m_{V_1}$):
\begin{equation}
\Gamma = \frac{|g|^2\sqrt{m_V^2-4m_{V_1}^2}}{192\pi\, m_V^2 m_{V_1}^4}
\bigl(m_V^6-8m_V^4 m_{V_1}^2+28m_V^2 m_{V_1}^4-48m_{V_1}^6\bigr)
\label{eq:VVV-equal}
\end{equation}
\end{agentresponse}

\section{Task 1 agent transcript}
\label{app:task1_transcript}

This appendix contains curated excerpts from the Claude Code session that completed Task~1.
The excerpts are organized into the four phases described in \cref{sec:task1_workflow}; the complete unedited session log is available in the data release (\cref{sec:data_availability}).


\subsection*{Phase 1: Enumeration and code generation}

\begin{agentuser}
Enumerate and compute the symbolic partial decay width $\Gamma(A \to BC)$ for every tree-level, Lorentz-invariant, single-vertex $1 \to 2$ process in four spacetime dimensions. Systematically sweep over parent spins $\{0, \frac{1}{2}, 1\}$ and all kinematically allowed daughter spin assignments. Retain complex couplings and distinct daughter masses throughout; equal-mass, massless, and real-coupling limits are taken as special cases. For each vertex type that admits both a vector--axial $(g_V, g_A)$ or scalar--pseudoscalar $(g_S, g_P)$ parameterization and a chiral $(g_L, g_R)$ parameterization, compute both in parallel and verify algebraic consistency under the basis change.

Validate a representative subset of the computed formulas against known Standard Model partial widths, using known masses and couplings.

Deliverable: a complete reference table of independent decay-width formulas in both coupling bases where applicable, together with useful limits, chiral cross-checks, NDA cross-checks, and Standard Model validation. Include figures that make the physical content of the catalog accessible and intuitive --- the key patterns and threshold structure should be assessable from the figures alone. Finally, provide a summary of phenomenological insights from the analysis.
\end{agentuser}

\begin{agentresponse}
This is a comprehensive catalog computation. Let me plan and execute systematically.
\end{agentresponse}

\begin{agentresponse}
I'll systematically enumerate all single-vertex 1$\to$2 decays across spins \{0, 1/2, 1\}, compute symbolic widths in both coupling bases where applicable, and build the complete catalog. Let me start.

\textbf{Spin-0 parent}: $S \to f \bar{f}$ (SFF vertices)
\end{agentresponse}

\begin{agenttool}{ComputeSymbolicAmplitude}
\small\textbf{Input:}
\begin{lstlisting}[style=jsonstyle]
"diagram": {
  "initial": [{"label": "S", "spin": 0}],
  "final": [{"label": "f", "spin": "1/2"},
            {"label": "fbar", "spin": "1/2"}],
  "vertices": [{"type": "scalar-va"}]
},
"script_name": "S_to_ff_scalarva",
"assume_real_couplings": false
\end{lstlisting}
\small\textbf{Result:}
\begin{lstlisting}[style=jsonstyle]
"status": "ok",
"process_type": "DECAY_1TO2",
"script_path": "S_to_ff_scalarva.wl"
\end{lstlisting}
\end{agenttool}

\begin{agenttool}{ComputeSymbolicAmplitude}
\small \texttt{script\_name=S\_to\_ff\_chiral}
\end{agenttool}

\begin{agentresponse}
\textbf{Spin-0 parent}: $S \to S_{1} S_{2}$ (SSS)
\end{agentresponse}

\begin{agenttool}{ComputeSymbolicAmplitude}
\small \texttt{script\_name=S\_to\_S1S2\_SSS}
\end{agenttool}

\begin{agentresponse}
\textbf{Spin-0 parent}: $S \to V_{1} V_{2}$ (SVV variants)
\end{agentresponse}

\begin{agenttool}{ComputeSymbolicAmplitude $(\times 3)$}
\begin{enumerate}[leftmargin=1.5em, itemsep=1pt, parsep=0pt]
  \item \small \texttt{script\_name=S\_to\_V1V2\_SVV}
  \item \small \texttt{script\_name=S\_to\_V1V2\_dim5FF}
  \item \small \texttt{script\_name=S\_to\_V1V2\_dim5FFdual}
\end{enumerate}
\end{agenttool}

\begin{agentresponse}
\textbf{Spin-1/2 parent}: $f_{1} \to S \bar{f}_{2}$ (SFF vertices)
\end{agentresponse}

\begin{agenttool}{ComputeSymbolicAmplitude $(\times 2)$}
\begin{enumerate}[leftmargin=1.5em, itemsep=1pt, parsep=0pt]
  \item \small \texttt{script\_name=f\_to\_Sf\_scalarva}
  \item \small \texttt{script\_name=f\_to\_Sf\_chiral}
\end{enumerate}
\end{agenttool}

\begin{agentresponse}
\textbf{Spin-1/2 parent}: $f_{1} \to V \bar{f}_{2}$ (VFF vertices)
\end{agentresponse}

\begin{agenttool}{ComputeSymbolicAmplitude $(\times 4)$}
\begin{enumerate}[leftmargin=1.5em, itemsep=1pt, parsep=0pt]
  \item \small \texttt{script\_name=f\_to\_Vf\_va}
  \item \small \texttt{script\_name=f\_to\_Vf\_chiral}
  \item \small \texttt{script\_name=f\_to\_Vf\_tensor}
  \item \small \texttt{script\_name=f\_to\_Vf\_tensorchiral}
\end{enumerate}
\end{agenttool}

\begin{agentresponse}
\textbf{Spin-1 parent}: $V \to f \bar{f}$ (VFF vertices)
\end{agentresponse}

\begin{agenttool}{ComputeSymbolicAmplitude $(\times 4)$}
\begin{enumerate}[leftmargin=1.5em, itemsep=1pt, parsep=0pt]
  \item \small \texttt{script\_name=V\_to\_ff\_va}
  \item \small \texttt{script\_name=V\_to\_ff\_chiral}
  \item \small \texttt{script\_name=V\_to\_ff\_tensor}
  \item \small \texttt{script\_name=V\_to\_ff\_tensorchiral}
\end{enumerate}
\end{agenttool}

\begin{agentresponse}
\textbf{Spin-1 parent}: $V \to S_{1} S_{2}$ (SSV), $V \to S V'$ (SVV), $V \to V_{1} V_{2}$ (VVV)
\end{agentresponse}

\begin{agenttool}{ComputeSymbolicAmplitude $(\times 3)$}
\begin{enumerate}[leftmargin=1.5em, itemsep=1pt, parsep=0pt]
  \item \small \texttt{script\_name=V\_to\_S1S2\_SSV}
  \item \small \texttt{script\_name=V\_to\_SV\_SVV}
  \item \small \texttt{script\_name=V\_to\_V1V2\_VVV}
\end{enumerate}
\end{agenttool}

\begin{agentresponse}
All 19 symbolic amplitude scripts generated. Now executing all in batch.
\end{agentresponse}

\subsection*{Phase 2: Batch execution}

\begin{agenttool}{RunWolframScriptBatch}
\small\textbf{Input:}
\begin{lstlisting}[style=jsonstyle]
"script_paths": [
  "S_to_ff_scalarva.wl", "S_to_ff_chiral.wl",
  "S_to_S1S2_SSS.wl", "S_to_V1V2_SVV.wl",
  "S_to_V1V2_dim5FF.wl", "S_to_V1V2_dim5FFdual.wl",
  "f_to_Sf_scalarva.wl", "f_to_Sf_chiral.wl",
  "f_to_Vf_va.wl", "f_to_Vf_chiral.wl",
  "f_to_Vf_tensor.wl", "f_to_Vf_tensorchiral.wl",
  "V_to_ff_va.wl", "V_to_ff_chiral.wl",
  "V_to_ff_tensor.wl", "V_to_ff_tensorchiral.wl",
  "V_to_S1S2_SSV.wl", "V_to_SV_SVV.wl",
  "V_to_V1V2_VVV.wl"
],
"max_concurrent": 4, "timeout": 180
\end{lstlisting}
\small\textbf{Result:}
\begin{lstlisting}[style=jsonstyle]
"total_scripts": 19, "succeeded": 18, "failed": 1,
"total_time_s": 192.9
\end{lstlisting}
\end{agenttool}

\begin{agentresponse}
18/19 succeeded. $V \to f \bar{f}$ (tensor) timed out. Let me retry with longer timeout and with in-script simplifications to help Mathematica.
\end{agentresponse}

\begin{agenttool}{ComputeSymbolicAmplitude | RunWolframScript}
\small \texttt{script\_name=V\_to\_ff\_tensor\_v2, timeout=300} $\to$ \texttt{success: true}
\end{agenttool}

\subsection*{Phase 3: Validation}

\begin{agentresponse}
All 19 formulas computed. Now I'll verify chiral $\leftrightarrow$ VA consistency and compute key limits. Let me run the cross-checks and limits in a structured way.
\end{agentresponse}

\begin{agenttool}{RunWolframScript}
\small \texttt{6 chiral $\leftrightarrow$ VA cross-checks} $\to$ \texttt{all zeros}
\end{agenttool}

\begin{agentresponse}
All 6 chiral $\leftrightarrow$ VA cross-checks pass (all zeros). Now computing key limits.
\end{agentresponse}

\begin{agenttool}{SimplifyResultBatch}
\small 17 special-case limits (equal-mass, massless) $\to$ 15/17 succeeded; 2 VA massless limits timed out
\end{agenttool}

\begin{agentresponse}
15/17 limits computed. The two VA massless limits are easily derived from the chiral results by basis change: $S \to ff$ gives $m_S(|g_S|^2+|g_P|^2)/(8\pi)$ and $V \to ff$ gives $m_V(|g_V|^2+|g_A|^2)/(12\pi)$.

Now let me run SM validation, NDA cross-checks, and fetch PDG data in parallel.
\end{agentresponse}

\begin{agenttool}{PDGDatabase $(\times 4)$}
\begin{enumerate}[leftmargin=1.5em, itemsep=1pt, parsep=0pt]
  \item \small \texttt{particle=Higgs, property=all}
  \item \small \texttt{particle=W boson, property=all}
  \item \small \texttt{particle=Z, property=all}
  \item \small \texttt{particle=top, property=all}
\end{enumerate}
\end{agenttool}

\begin{agenttool}{EstimateDecayWidthNDA $(\times 5)$}
\begin{enumerate}[leftmargin=1.5em, itemsep=1pt, parsep=0pt]
  \item \small \texttt{H $\to$ b$\bar{\texttt{b}}$} $\to$ $\Gamma_{\text{NDA}} = 4.31\times 10^{-3}$~GeV
  \item \small \texttt{Z $\to$ e$^+$e$^-$} $\to$ $\Gamma_{\text{NDA}} = 88.5\times 10^{-3}$~GeV
  \item \small \texttt{W $\to$ e$\nu$} $\to$ $\Gamma_{\text{NDA}} = 227.0\times 10^{-3}$~GeV
  \item \small \texttt{t $\to$ Wb} $\to$ \textcolor{red}{\texttt{Diagram Validation Failed}} (first attempt)
  \item \small \texttt{t $\to$ Wb} (corrected) $\to$ $\Gamma_{\text{NDA}} = 1.386$~GeV
\end{enumerate}
\end{agenttool}

\begin{agentresponse}
SM validation is excellent:
\begin{itemize}
  \item $Z$$\to$$e^+ e^-$: 83.91 vs 83.93 MeV (PDG) --- 0.02\% agreement
  \item $W$$\to e \nu$: 227.2 vs 228.9 MeV --- 0.7\%
  \item $t$$\to$$Wb$: 1.479 vs 1.424 GeV --- 3.8\%
  \item $H$$\to$$b$$\bar{b}$ (running mass): 1.91 vs 1.96 MeV --- 2.6\%
\end{itemize}
Now let me create the publication-quality figures.
\end{agentresponse}

\subsection*{Phase 4: Synthesis}

\begin{agenttool}{Write $(\times 4)$}
\begin{enumerate}[leftmargin=1.5em, itemsep=1pt, parsep=0pt]
  \item \small \texttt{fig\_threshold.py} --- threshold power-law structure
  \item \small \texttt{fig\_svv\_operators.py} --- SVV operator comparison
  \item \small \texttt{fig\_longitudinal.py} --- longitudinal polarization enhancement
  \item \small \texttt{fig\_sm\_validation.py} --- SM validation and top width anatomy
\end{enumerate}
\end{agenttool}

\begin{agenttool}{Bash $(\times 4)$}
\small Execute all figure scripts $\to$ 4 PDFs generated
\end{agenttool}

\begin{agentresponse}
The left panel has a scaling issue --- the longitudinal enhancement in SVV/VVV channels overwhelms the plot. Let me redesign the figures to properly reveal the physics.
\end{agentresponse}

\begin{agenttool}{Write $(\times 4)$}
\small Revised figure scripts with separate panels for threshold, SVV operators, longitudinal enhancement, and SM validation
\end{agenttool}

\begin{agenttool}{Bash | Read $(\times 4)$}
\small Regenerate and inspect all four figures
\end{agenttool}

\begin{agentresponse}
The $H \to b \bar{b}$ NDA bar overflows because NDA used pole mass while tree-level used running mass. Let me fix this for consistency.
\end{agentresponse}

\begin{agenttool}{Edit | Bash | Read}
\small Fix mass inputs in SM validation figure $\to$ regenerate
\end{agenttool}

\begin{agentresponse}
All four figures look publication-quality. Now let me write the comprehensive LaTeX document.
\end{agentresponse}

\begin{agenttool}{Write}
\small \texttt{decay\_catalog.tex} --- 12-page reference document
\end{agenttool}

\begin{agenttool}{Bash $(\times 2)$}
\small \texttt{pdflatex} (two passes) $\to$ \texttt{decay\_catalog.pdf}
\end{agenttool}

\begin{agentresponse}
The catalog is complete. Here's a summary of everything that was computed and delivered:
\vspace{0.1in}

\textbf{Complete Catalog: 19 Independent Decay Width Formulas}\vspace{0.05in}

\begin{tabular}{llll}
\toprule
Parent & Final state & Vertex types & Formulas \\
\midrule
Spin-0 & $f\bar{f}$ & scalar-VA, chiral & 2 \\
Spin-0 & $S_1 S_2$ & SSS & 1 \\
Spin-0 & $V_1 V_2$ & SVV, $\phi FF$, $\phi F\tilde{F}$ & 3 \\
Spin-1/2 & $S\bar{f}$ & scalar-VA, chiral & 2 \\
Spin-1/2 & $V\bar{f}$ & VA, chiral, tensor, tensor-chiral & 4 \\
Spin-1 & $f\bar{f}$ & VA, chiral, tensor, tensor-chiral & 4 \\
Spin-1 & $S_1 S_2$ & SSV & 1 \\
Spin-1 & $SV'$ & SVV & 1 \\
Spin-1 & $V_1 V_2$ & VVV & 1 \\
\bottomrule
\end{tabular}

\vspace{0.1in}
\textbf{Validation:} 6/6 chiral cross-checks pass; SM benchmarks agree within 0.02--3.8\% of PDG; NDA within $\sim$6\% of exact.

\vspace{0.1in}
\textbf{Deliverables:} 12-page LaTeX reference document, 4 publication-quality figures, Python module encoding all 19 formulas, 19 FeynCalc scripts with JSON sidecars.
\end{agentresponse}

\section{Task 2 agent transcript}
\label{app:task2_transcript}

This appendix contains curated excerpts from the Claude Code session that completed Task~2.
The excerpts are organized into the four phases described in \cref{sec:task2_workflow}; the complete unedited session log is available in the data release (\cref{sec:data_availability}).


\subsection*{Phase 1: Experimental context and diagram enumeration}

\begin{agentuser}
Determine the largest number of $e^+e^-$ pairs $n$ for which the SM decay $\mu^+ \to \bar{\nu}_\mu \nu_e + n(e^+e^-) + e^+$ remains observable at current or planned muon experiments. Begin by establishing the experimental landscape: existing measurements, planned experiments (Mu3e, etc.), and their sensitivity reaches. For each multiplicity, enumerate the contributing tree-level diagrams and estimate the branching ratio using NDA. Continue increasing $n$ until the rate falls below experimental sensitivity.

Cross-check the dominant diagram class at the lowest and next-to-lowest non-trivial multiplicity against an exact MadGraph calculation, using coupling order restrictions to isolate the relevant diagrams. Provide a brief analytical estimate of the per-pair suppression factor and compare it with the observed numerical scaling.

Deliverable: a comprehensive table of diagram counts, NDA widths, scalings, and branching ratios for each final-state multiplicity, identifying the experimental frontier. Include ``diagram metadata'' figures that make the enumeration and NDA results intuitive. Additionally create a summary figure illustrating how the total branching ratio scales with $n$ relative to experimental sensitivity thresholds. Conclude with a discussion of the physics driving the multiplicity scaling and its implications for current and planned experiments.
\end{agentuser}

\begin{agentresponse}
I'll tackle this systematically: set up the workspace, enumerate diagrams for increasing multiplicity, estimate branching ratios, and cross-check with MadGraph. Let me start by fetching the tools I need and setting up directories.
\end{agentresponse}

\begin{agentresponse}
Let me start by gathering experimental context, PDG data, and enumerating diagrams for the first multiplicities in parallel.
\end{agentresponse}

\begin{agenttool}{PDGSearch}
\small\textbf{Input:} \texttt{"query": "muon"}
\;\textbf{Result:} $m_\mu = 0.10566$ GeV
\end{agenttool}

\begin{agentsubagent}{Research muon experiment sensitivity}
\small 35 web searches $\to$ Established SINDRUM measurement BR$(n\!=\!1) = (3.4 \pm 0.4)\times10^{-5}$, Mu3e Phase~I SES $= 2\times10^{-15}$, Phase~II SES $\sim 10^{-16}$, HiMB $\sim 10^{-18}$. Published $n=2$ prediction BR $= (3.93 \pm 0.01)\times10^{-10}$ from Hostert \textit{et al.}
\end{agentsubagent}

\begin{agenttool}{EnumerateDiagrams}
\small\textbf{Input:}
\begin{lstlisting}[style=jsonstyle]
"initial": ["mu+"],
"final": ["e+", "nu_e", "nu_mubar"],
"output_format": "tikz"
\end{lstlisting}
\small\textbf{Result:}
\begin{lstlisting}[style=jsonstyle]
"n_diagrams": 1,
"process": "mu+ -> e+ nu_e nu_mubar"
\end{lstlisting}
\end{agenttool}

\begin{agenttool}{EnumerateDiagrams}
\small \texttt{final=["e+","e+","e-","nu\_e","nu\_mubar"]} $\to$ \texttt{n\_diagrams: 18}
\end{agenttool}

\begin{agenttool}{EstimatePhaseSpace $(\times 4)$}
\begin{enumerate}[leftmargin=1.5em, itemsep=1pt, parsep=0pt]
  \item \small $n_{\text{body}}=3$ $\to$ $\Phi_3 = 1.41\times10^{-6}$
  \item \small $n_{\text{body}}=5$ $\to$ $\Phi_5 = 9.76\times10^{-17}$
  \item \small $n_{\text{body}}=7$ $\to$ $\Phi_7 = 8.13\times10^{-28}$
  \item \small $n_{\text{body}}=9$ $\to$ $\Phi_9 = 1.73\times10^{-39}$
\end{enumerate}
\end{agenttool}

\begin{agentresponse}
Good initial results. Now let me read the diagram summaries, run NDA estimates, and enumerate $n=2$ in parallel.
\end{agentresponse}

\begin{agenttool}{Read $(\times 2)$}
\small Diagram summary tables for $n=0$ (1 diagram, 1 class) and $n=1$ (18 diagrams, 3 classes)
\end{agenttool}

\begin{agenttool}{EnumerateDiagrams}
\small \texttt{final=[7 particles]} $\to$ \texttt{n\_diagrams: 1122}
\end{agenttool}

\begin{agentresponse}
Excellent! The diagram counts are growing rapidly:
\begin{itemize}
  \item $n=0$: 1 diagram
  \item $n=1$: 18 diagrams (4 dominant with 1 heavy $W$)
  \item $n=2$: 1,122 diagrams (84 dominant with 1 heavy $W$)
\end{itemize}
Now let me read the $n=2$ summary, run NDA branching ratios, enumerate $n=3$, and start MadGraph cross-checks --- all in parallel.
\end{agentresponse}

\subsection*{Phase 2: NDA estimation}

\begin{agenttool}{EstimateBranchingRatioNDA}
\small\textbf{Input:}
\begin{lstlisting}[style=jsonstyle]
"diagram_classes": [
  {"diagram_path": "diagrams_.../heavy_1/",
   "n_diagrams": 1, "n_heavy": 1}
],
"reference_width": 2.996e-19,
"process_label": "mu+ -> e+nu_enu_mu (n=0)"
\end{lstlisting}
\small\textbf{Result:}
\begin{lstlisting}[style=jsonstyle]
"branching_ratio": 2.86,
"width_per_diagram": 8.56e-19,
"formula": "g_w^4 M^5 / (64 pi^3 M_W^4)"
\end{lstlisting}
\end{agenttool}

\begin{agenttool}{EstimateBranchingRatioNDA $(\times 2)$}
\begin{enumerate}[leftmargin=1.5em, itemsep=1pt, parsep=0pt]
  \item \small $n=1$: BR $= 1.85\times10^{-5}$ (3 classes, 18 diagrams)
  \item \small $n=2$: BR $= 1.89\times10^{-9}$ (5 classes, 1{,}122 diagrams)
\end{enumerate}
\end{agenttool}

\begin{agenttool}{EnumerateDiagrams}
\small \texttt{final=[9 particles]} $\to$ \texttt{n\_diagrams: 149400} (7 classes)
\end{agenttool}

\begin{agenttool}{Write $(\times 2)$}
\small MadGraph command cards: \texttt{mg5\_n1\_dominant.mg5} (QED$\leq$4) and \texttt{mg5\_n1\_full.mg5} (no restriction)
\end{agenttool}

\begin{agentresponse}
The $n=3$ data is in (149,400 diagrams!). Now let me run NDA for $n=3$, launch MadGraph cross-checks, and prepare for $n=4$ enumeration.
\end{agentresponse}

\subsection*{Phase 3: MadGraph cross-checks}

\begin{agenttool}{MadGraphFromRunCard}
\small\textbf{Input:}
\begin{lstlisting}[style=jsonstyle]
"command_card": "scripts/mg5_n1_dominant.mg5",
"data_dir": "scripts/mg5_runs"
\end{lstlisting}
\small\textbf{Result:}
\begin{lstlisting}[style=jsonstyle]
"width_gev": 1.066e-23,
"width_gev_error": 4.109e-25
\end{lstlisting}
\end{agenttool}

\begin{agenttool}{EstimateBranchingRatioNDA}
\small $n=3$: BR $= 8.26\times10^{-13}$ (7 classes, 149{,}400 diagrams)
\end{agenttool}

\begin{agenttool}{MadGraphFromRunCard}
\small \texttt{mg5\_n1\_full.mg5} $\to$ same result (confirming subleading classes negligible)
\end{agenttool}

\begin{agentresponse}
Key results so far:

\textbf{MadGraph $n=1$ cross-check}: $\Gamma = 1.066\times10^{-23}$ GeV $\to$ BR $= 3.56\times10^{-5}$ (vs experiment $3.4\times10^{-5}$). The QED$\leq$4 and full results are identical, confirming the 1-heavy-$W$ class dominates completely.

\textbf{NDA/MG comparison}: NDA gives $5.55\times10^{-24}$, MG gives $1.07\times10^{-23}$ $\to$ NDA off by factor $\sim$2, typical accuracy.

Now let me launch MadGraph for $n=2$, try $n=4$ enumeration, and start building figures.
\end{agentresponse}

\begin{agenttool}{MadGraphFromRunCard}
\small \texttt{mg5\_n2\_dominant.mg5} (QED$\leq$6) $\to$ $\Gamma = 1.301\times10^{-28}$ GeV
\end{agenttool}

\begin{agenttool}{EnumerateDiagrams}
\small \texttt{final=[11 particles]} $\to$ \textcolor{red}{\texttt{MCP error: Connection closed}} ($n=4$ too complex)
\end{agenttool}

\begin{agentresponse}
Excellent! \textbf{MadGraph $n=2$}: $\Gamma = 1.301\times10^{-28}$ GeV $\to$ BR $= 4.34\times10^{-10}$, matching the published value of $3.93\times10^{-10}$. The $n=4$ enumeration timed out (too many diagrams), which itself is informative.

Let me now read the representative TikZ diagrams and start building the figures and LaTeX document.
\end{agentresponse}

\subsection*{Phase 4: Synthesis}

\begin{agenttool}{Read $(\times 4)$}
\small Representative TikZ diagrams for $n=0$, $n=1$, $n=2$ dominant classes + \texttt{mplstyle}
\end{agenttool}

\begin{agenttool}{Write $(\times 2)$ | Bash $(\times 2)$}
\small \texttt{plot\_br\_scaling.py} and \texttt{plot\_diagram\_metadata.py} $\to$ 2 PDFs generated
\end{agenttool}

\begin{agentresponse}
Both figures look publication-quality. Now let me write the comprehensive LaTeX document and create one more figure showing the per-pair suppression analysis.
\end{agentresponse}

\begin{agenttool}{Write | Bash}
\small \texttt{plot\_suppression.py} $\to$ \texttt{suppression\_analysis.pdf}
\end{agenttool}

\begin{agenttool}{Write}
\small \texttt{multipair\_muon\_decay.tex} --- 8-page analysis document
\end{agenttool}

\begin{agenttool}{Bash $(\times 3)$}
\small \texttt{pdflatex} (three passes for cross-references) $\to$ \texttt{multipair\_muon\_decay.pdf}
\end{agenttool}

\begin{agentresponse}
The document is complete and publication-quality across all 8 pages. Let me clean up the workspace and update the findings summary.
\end{agentresponse}

\begin{agentresponse}
Here is a summary of the complete analysis:

\textbf{Experimental Frontier: $n = 3$}

\begin{tabular}{llllll}
\toprule
$n$ & Diagrams & Dominant (1$W$) & BR & Status \\
\midrule
0 & 1 & 1 & $\sim$1 & Measured \\
1 & 18 & 4 & $3.56\times10^{-5}$ & Measured (SINDRUM) \\
2 & 1,122 & 84 & $4.34\times10^{-10}$ & Mu3e Phase~I ($\sim$$10^6$ events) \\
3 & 149,400 & 4,320 & $\sim$$10^{-13}$ & Mu3e Phase~I ($\sim$$10^2$ events) \\
4 & $\sim$$10^7$ & $\sim$$3\times10^5$ & $\sim$$10^{-16}$ & Phase~II / HiMB (marginal) \\
\bottomrule
\end{tabular}

\vspace{0.1in}
\textbf{Key Physics:}
\begin{itemize}
  \item \textbf{Per-pair suppression}: $\sim$$10^{-5}$, driven by $(\alpha/\pi)^2 \approx 5\times10^{-6}$ modulated by rapidly growing diagram multiplicity
  \item \textbf{MadGraph cross-checks} at $n=1,2$ validate NDA to within a factor of 2--4 and match the SINDRUM measurement and published predictions
  \item \textbf{Dominant topology}: Single-$W$ exchange dressed by $n$ virtual photons converting to $e^+e^-$ pairs; subleading classes suppressed by $(m_\mu/M_W)^4 \sim 10^{-13}$ per additional heavy propagator
  \item \textbf{Interference transition}: Constructive at $n=1$ (NDA underestimates), destructive at $n=2$ (NDA overestimates), stabilizing the effective suppression
\end{itemize}

\textbf{Deliverables:} 8-page LaTeX document, 3 publication-quality figures, diagram enumerations for $n=0$--$3$ with TikZ visualizations.
\end{agentresponse}

\end{document}